\documentclass[conference]{IEEEtran}
\usepackage{cite}
\usepackage{amsmath,amssymb,amsfonts}
\usepackage{algorithmic}
\usepackage{graphicx}
\usepackage{textcomp}
\usepackage{xcolor}
\usepackage[hyphens]{url}
\usepackage{dblfloatfix}
\usepackage{siunitx}
\usepackage{enumitem}
\usepackage{soul}
\usepackage[caption=false]{subfig}
\usepackage[font=small,labelfont=bf]{caption}
\usepackage[section]{placeins} 
\usepackage{xspace}

\newcommand*\rot{\rotatebox{90}}
\newcommand*\OK{\ding{51}}
\newcommand{\trace}{trace}

\newcolumntype{P}[1]{>{\centering\arraybackslash}m{#1}}

\def\BibTeX{{\rm B\kern-.05em{\sc i\kern-.025em b}\kern-.08em
    T\kern-.1667em\lower.7ex\hbox{E}\kern-.125emX}}

\newcommand{\ignore}[1]{}
\newcommand{\myalgo}{Puppeteer\xspace}
\newcommand{\suitapmech}{Algorithm}

\title{Puppeteer: A Random Forest-based Manager for Hardware Prefetchers across the Memory Hierarchy \vspace{-0.2in}}
\author{
	Furkan Eris$^1$,
	Marcia~S~Louis$^1$,
	Kubra Cilingir$^1$,
	\normalfont {Jos\'e~L.~Abell\'an$^2$,
	Ajay~Joshi$^1$}\\
	\normalfont{\small $^1$ECE Department, Boston University;}
	\normalfont{\small $^2$CS Department, UCAM;}\\
	\normalfont{\small \{fe, marcia93, kubra joshi\}@bu.edu,}
  \normalfont{\small jlabellan@ucam.edu}
  \vspace{-5mm}
  }

\begin{document}
\maketitle
\thispagestyle{plain}
\pagestyle{plain}

\begin{abstract}
Over the years, processor throughput has steadily increased.
However, the memory throughput has not increased at the same rate, which has led to the memory wall problem in turn increasing the gap between effective and theoretical peak processor performance.
To cope with this, there has been an abundance of work in the area of data/instruction prefetcher designs.
Broadly, prefetchers predict future data/instruction address accesses and proactively fetch data/instructions in the memory hierarchy with the goal of lowering data/instruction access latency.
To this end, one or more prefetchers are deployed at each level of the memory hierarchy, but typically, each prefetcher gets designed in isolation without comprehensively accounting for other prefetchers in the system.
As a result, individual prefetchers do not always complement each other, and that leads to lower average performance gains and/or many negative outliers. In this work, we propose~\myalgo, which is a hardware prefetcher manager that uses a suite of random forest regressors to determine at runtime which prefetcher should be ON at each level in the memory hierarchy, such that the prefetchers complement each other and we reduce the data/instruction access latency.
Compared to a design with no prefetchers, using \myalgo~we improve IPC by 46.0\% in 1 Core (1C), 25.8\% in 4 Core (4C), and 11.9\% in 8 Core (8C) processors on average across traces generated from SPEC2017, SPEC2006, and Cloud suites with $\sim$10KB overhead.
Moreover, we also reduce the number of negative outliers by over 89\%, and the performance loss of the worst-case negative outlier from 25\% to only 5\% compared to the state-of-the-art.
\end{abstract}

\section{INTRODUCTION}
\label{sec:intr}

Instruction and data prefetching \cite{falsafi2014primer} are commonly used in today's processors to overcome the memory wall problem~\cite{wulf1995hitting}.
The key idea behind prefetching is identifying the current memory access pattern and predicting addresses to proactively fetch instructions and data into the cache to avoid cache misses.
Prefetching hides the large memory access latency, and in turn, improves processor performance.
As a result, modern processors employ multiple prefetchers to cover a wide range of applications.
Consequently, existing prefetchers do not improve the performance of all applications; in some cases they hurt application performance by prefetching the wrong memory addresses \cite{lee2012prefetching}.
These incorrect prefetches use up the precious memory bandwidth and the limited space in the cache hierarchy.
This increases data and instruction access latency, which hurts application performance.


\ignore{
The hardware prefetching system in today's processors consists of multiple prefetchers, and this prefetcher system follows a pre-determined protocol for prefetcher priority, prefetcher throttling, etc. to regulate the interactions between prefetchers. Generally, this protocol is tuned for the common case and this tuning is done at design time in both AMD and Intel CPUs \cite{bios2010kernel,guide2011intel}.} 
\ignore{
{\color{black}Only tuning for the common case can cause complex interactions between prefetchers which can lead to IPC losses.} Despite this tuning,}

To evaluate a prefetcher's performance, we can use scope and accuracy as the metrics \cite{bhatia2019perceptron, kondguli2018division}. 
A prefetcher with high prefetching accuracy usually has limited scope, i.e., it is very good at identifying a limited number of memory access patterns and can accurately prefetch data/instructions if those specific memory access patterns exist.
However, such a prefetcher fails to identify other memory access patterns.
Conversely, a prefetcher with broad scope caters to a wide variety of memory access patterns, but it has low accuracy. 

\ignore{
{\color{black}Each prefetcher tracks a specific type of traffic. For example, a stride prefetcher tracks strided accesses for a single stride, and it uses thresholds tuned for the average strided sequences. For applications that do not have strided accesses, this stride prefetcher may be suboptimal, leading to cache thrashing. In a prefetching system consisting of multiple prefetchers, this issue is more pronounced because the architectural resources are shared among all prefetchers;}}
\ignore{
\textbf{Duplicate Scope -} {\color{black} Given that prefetchers in a prefetcher system are trained independently to track a specific type of traffic, multiple prefetchers may trigger prefetch requests for the same memory patterns. This leads to wasted bandwidth and increased power consumption.} \break

{\color{black} and 2) Different prefetchers latch onto memory access patterns at different speeds, and so a prefetcher's behavior over time can be affected by the traffic of the other prefetchers. These differences in temporal behavior can cause faulty synchronization among prefetchers. Variations in the accuracy of each prefetcher and faulty synchronization can lead to a drop in performance.}}

Effectively, we need to find a balance between the accuracy and scope of the prefetcher.
One way to balance scope and accuracy is to use multiple high accuracy prefetchers at each level of the memory hierarchy, as is the case in AMD and Intel processors~\cite{Inteli9,AMDryzen2017,bios2010kernel, guide2011intel}.
Each prefetcher is customized to identify a specific type of memory access pattern and make a prefetching prediction.
However, having multiple prefetchers operating at each level in the memory hierarchy can lead to the following issues:
\begin{itemize}[leftmargin=*]
    \item Given that each prefetcher is trained independently to track a specific type of traffic and simultaneously share microarchitectural resources, prefetchers can sabotage each other during runtime. 
    A prefetcher may trigger prefetch requests that evict cache lines that another prefetcher has accurately prefetched.
    This behavior leads to a loss in performance, wasted memory access bandwidth, and increased power consumption.
    \item Different prefetchers (either at the same level or across levels in the memory hierarchy) latch onto memory access patterns at different speeds. So a prefetcher's prediction can be influenced by the traffic generated by other prefetchers. 
    These differences in temporal behavior can cause faulty synchronization among prefetchers and lead to a drop in application performance.
\end{itemize}

Essentially, prefetchers compete for resources, and at times, sabotage each other. 
To validate our argument, we use traces\footnote{A \trace~is a group of instructions that represent a specific behavior. One or more \trace s can be used to represent the behavior of a benchmark. 
For example, a benchmark with consistent looping behavior can be represented by one \trace~corresponding to a single iteration of the loop. 
One or more unique representative \trace s are generated from each benchmark
\cite{perelman2003picking}.} generated from SPEC2017, SPEC2006, and Cloud benchmark suites and run these traces on an OoO processor that uses a different prefetcher at each level of the memory hierarchy (details of the particular evaluation methodology in Section~\ref{sec:methodology}).
We execute these traces using two state-of-the-art prefetcher solutions -- IPCP~\cite{Pakalapati2020ipcp} and EIP~\cite{ros2020entangling}, who were the winners of prior prefetching competitions \cite{ipc1, dpc3, champsim}. These prefetcher solutions use different prefetchers at each level of the memory hierarchy, i.e., different prefetcher system configurations (PSC)\footnote{A PSC specifies which prefetcher is switched ON at each level of the memory in the system. We denote a PSC using the following format \texttt{<prefetcher-in-L1I\$>-<prefetcher-in-L1D\$>}
\texttt{-<prefetcher-in-L2\$>-<prefetcher-in-LLC\$>}.}.
Both prefetcher solutions improve performance compared to the no prefetching case. 
However, IPCP (\textit{no-ipcp-ipcp-nl}) (further details on the types of prefetchers are given in Table \ref{tab:evaluated-solutions}) shows better performance over EIP in 100 out of 232 traces with the largest performance gain of 56\% in \texttt{602.gcc\_s-2226B}.
EIP (\textit{EIP-nl-spp-no}) shows better performance over IPCP in the remaining 132 traces and has the largest performance gain of 89\% in \texttt{server\_036}.

\begin{figure}[t] 
  \centering
  \includegraphics[width=\columnwidth]{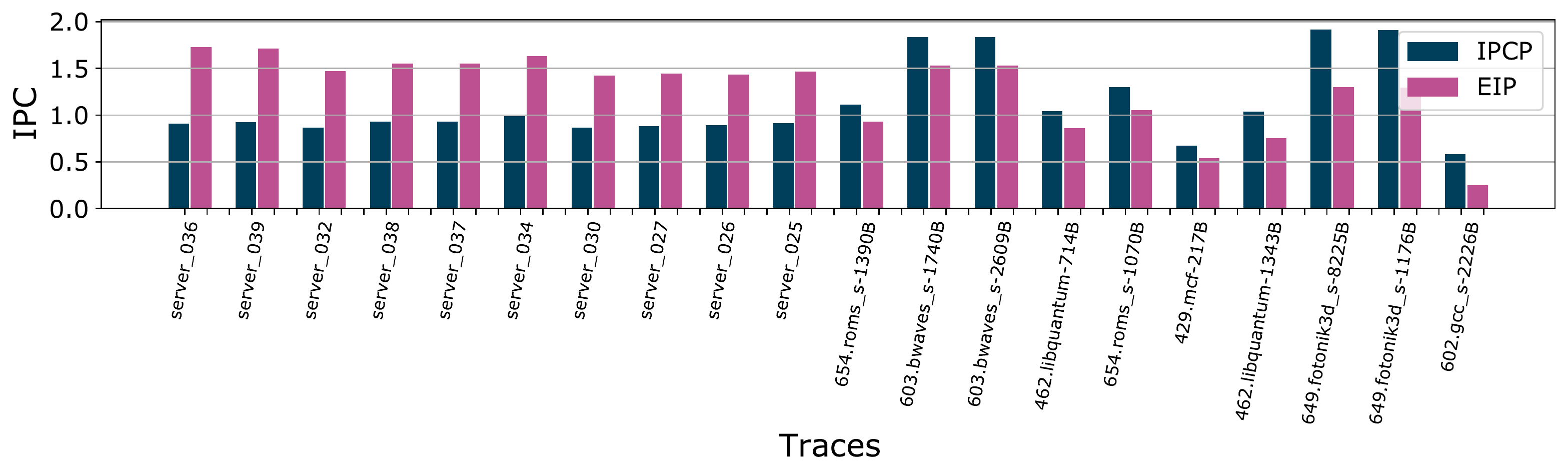}
  \vspace{-0.2in}
  \caption{\textbf{Motivational Example -} Performance of a processor when using IPCP and EIP prefetchers. Here we show the 20 traces with the highest difference in performance out of 232 traces that we evaluated.}
  \label{fig:motivation}
  \vspace{-0.25in}
\end{figure}

In Figure \ref{fig:motivation}, we show a subset of traces that have a significant difference in their IPC when using IPCP and EIP prefetcher solutions.
The standard methodology would have us select EIP because it has on average 7.8\% performance gain over IPCP, but that would come at the cost of having a performance loss for 100 out of the 232 traces.

One way to address this problem is to have multiple different prefetchers at each cache level and switch ON a prefetcher based on the current program phase.
%
For example, one prior work that has attempted to leverage multiple prefetchers in the \textit{same} level of the memory hierarchy is by Kondugli et al.~\cite{kondguli2018division}. 
The authors propose a `composite prefetcher', which uses a priority queue as the control algorithm to select a prefetcher at a single level in the memory hierarchy. 
While this approach provides benefits, the  `composite prefetcher' priority queue is designed offline for a given set of applications.
For previously unseen applications, we will not necessarily see any performance improvement.
Furthermore, the control algorithm will not scale well as we increase the number of prefetchers in the system and target \textit{different} levels in the memory hierarchy.

What is really needed is a \textit{manager} that can successfully manage multiple prefetchers at each level in the memory hierarchy and has a low overhead.
This manager will choose the PSC that is suitable for the current phase of an application.
The chosen PSC should have prefetchers that complement each other for the current phase, reduce memory access overhead, and in turn improve application performance.
Given that multiple prefetchers are available at each level in the memory hierarchy, this `manager' will effectively determine which prefetcher should be ON/OFF at each level in the memory hierarchy, both across different phases of an application and across applications.
\ignore{
Contrary to the prior work \cite{liao2009machine, zeng2017long}, in our approach, we train our ML model to increase the overall processor IPC instead of training the model to improve the accuracy. 
We adapt a multi-label classification approach for prefetcher adaptation, where we leverage the implicit ranking among PSCs for each application to train our ML model to catch performance outliers. 
We train a unique random forest regressor per PSC to create a suite of random forest regressors. A more detailed discussion about our choice of multi-label classification and suite of random forests is provided in Section~\ref{sec:suitap}. {\color{black}We use PSC-invariant events as our features}. 
In particular, we choose events that are not affected by the choice of PSCs. We design our ML model with the hardware overhead in mind and aim to maximize IPC with minimum overhead.
\footnote{We would like to emphasize the fact that we are not proposing a novel prefetcher that uses machine learning. Jointly designing a machine learning-based manager that selects from an array of machine learning-based prefetchers is the logical next step in our work.}}
\textbf{In this paper we propose a machine learning (ML)-based hardware manager called \myalgo~that selects the PSC at runtime.}
Contrary to the prior work \cite{liao2009machine, zeng2017long, rahman2015maximizing} that focuses on training the ML model to improve the prefetch address prediction accuracy of the ML model, we train the ML model of \myalgo~to increase the overall system performance (quantified as IPC). 
Using our unique training strategy, we are able to specifically target training for application phases where the swing in IPC is much higher than in other regions. 
Thereby, we tailor \myalgo~for these phases and achieve high targeted-application-phase accuracy instead of just high overall model prediction accuracy. 

To manage the prefetchers across multiple cache levels at runtime, we propose a multi-regression ML-based approach.
We use the observed IPC of the various PSCs for different phases of each application to train our ML model. 
We train a unique random forest regressor per PSC (in our case, we have 5 different PSCs, which we pruned down from the 300 possible PSCs -- more details about this in Section \ref{sec:methodology}), to create a suite of random forests regressors. 
For features used in the ML model, we use events that can be tracked using hardware performance counters and whose behavior does not change with the choice of PSC, i.e., PSC-invariant events.
An example of such an event is the number of branch instructions in an application.
The branch instruction count does not change with the choice of PSC.
Using only PSC-invariant events, we can limit the number of executions per trace we must account for during training, making it easier to train the ML model in \myalgo (more details about the training approach are in in Section \ref{sec:training_details}).
\ignore{We design the ML model of \myalgo~to maximize IPC and minimize the area overhead required by the ML model.}
In summary, the contributions of our work are as follows:
\begin{itemize}[leftmargin=*]
\item We propose a novel ML-based runtime hardware manager called \myalgo~to manage the various prefetchers across the memory hierarchy to improve processor performance.
\item We design \myalgo~to use a set of PSC-invariant events (which can be tracked using hardware performance counters) as inputs and predict the PSC for the next instruction window\footnote{An instruction window is group of
instructions that are executed sequentially at runtime. 
We experimented with different instruction window sizes and did not see a large change in the performance of \myalgo. 
We set the size to 100,000 instructions.}. 
We co-optimize the hardware design and the ML model of \myalgo~with the goal of maximizing the overall application performance while minimizing the area overhead.
At runtime, at the end of an instruction window, \myalgo~predicts the IPC for each PSC and selects the PSC with the highest predicted IPC for the succeeding instruction window.
\item We train \myalgo~to maximize processor performance instead of the prefetch address prediction accuracy. We use only 10\% of the data for training to prevent overfitting and design \myalgo~with a low hardware overhead ($\sim$10KB). For the 232 traces we experiment with, \myalgo~achieves an average performance gain of 46.0\% in 1 Core (1C), 25.8\% in 4 Core (4C), and 11.9\% in 8 Core (8C) processors on average compared to a system with no prefetching. Moreover, we ensure \myalgo~reduces  negative outliers. When using \myalgo, we observe an 89\% reduction in the number of outliers, down to 8 outliers from 53, and a 20\% reduction in the IPC loss of the worst-case outlier compared to the state-of-the-art prior prefetcher managers. 
\item Finally, to the best of our knowledge, \myalgo~is the only manager that targets all cache levels in the memory hierarchy.
\end{itemize}

\section{Background and Related Work}
\label{sec:related}
\begin{table}[t]
\vspace{0.1in}
  \caption{\textbf{Overview of the Related Work on Prefetcher Managers -} We use \textit{NA}
  when \color{black}the paper lacks information about the respective metric. In the last column, `software' indicates the software prefetcher. For IPC gains, we report average IPC gain in 1C. 
  }
  \label{tab:prior}
\scriptsize
\centering
\begin{tabular}{P{0.9cm}|P{0.2cm}|P{1cm}|P{1.5cm}|P{1.5cm}|P{0.9cm}|P{0.9cm}}
\hline
\textbf{Work} & \textbf{\# Pf} & \textbf{\# Cache Levels Managed} & \textbf{IPC
Gain Over Baseline PSC} & \textbf{IPC Gain Over No Prefetching} 
&
\textbf{Heuristic or ML} & \textbf{Overhead} \\
\hline
\hline
\cite{rogers2019effects}~ &  1 & 1 & \multicolumn{2}{c|}{Only reports accuracy} & Heuristic & Software \\
\hline
\cite{kondguli2018division}~   & 1 & 1  & 5\% & 41\% & Heuristic &  $\sim4.6$KB \\ 
\hline
\cite{jimenez2012making}~ & 1 & 1 & 7.8\% (6\% from 1 application) & \textit{NA} & Heuristic & Software\\
\hline
\cite{kang2013hardware}~  & 0 & 1 & 11\% & No Pf baseline &  Heuristic &  Software \\
\hline 
\cite{liao2009machine} & 4 & 2 & \multicolumn{2}{c|}{Only reports ML accuracy} & ML & Software \\
\hline
\cite{rahman2015maximizing}~  & 1 & 2 & \multicolumn{2}{c|}{Only reports ML accuracy} & ML &  Software \\
\hline
\cite{bhatia2019perceptron} & 1 & 1 Pf targets 2 levels & 2.27\%  &
15.24\% &  ML & $\sim39$KB\\
\hline
\cite{hiebel2019machine} & 4 & 1 & -1\% to 3.6\% & -1\% to 8.5\% \ignore{(with one prefetcher disabled)} & ML &
Software\\
\hline
\cite{maldikar2014adaptive}~  & 1 & 1  & -5\% to 1\% & 23\% &  ML & Software \\
\hline
\textbf{\myalgo}~ & \textbf{7} & \textbf{All (4)} & \textbf{14.7\%} & \textbf{46\%} & 
\textbf{ML} & \textbf{$\sim10$KB}\\
\hline
\end{tabular}
\vspace{-0.2in}
\end{table} 

\subsection{Heuristic-based Prefetcher Managers}    
When a processor executes an application, there is a diverse set of interactions among the compute, memory and communication components of a computing system. 
These interactions are dependent on the processor (micro)architecture and the application, thus effectively making the processor a big finite state machine with an extremely large state space.
This makes it difficult to use deterministic techniques, which consider all possible states, for hardware management. 
As a result, over the past 20-30 years, heuristic algorithms have been used for hardware management, including for prefetcher management such as static priority queue-based approaches \cite{kondguli2018division, rogers2019effects} and rule-based prefetcher throttling approaches \cite{ebrahimi2009coordinated, ebrahimi2009techniques, ebrahimi2010fairness, ebrahimi2011prefetch, heirman2018near, hur2006memory, srinath2007feedback, liu2016thread}. 
These algorithms have low memory/area overhead and improve processor performance for the average case. 
However, with the ever-increasing complexity of processors and the diversity of applications, heuristic algorithms are no longer effective.
Heuristic algorithms are extremely dependent on the hardware/operating conditions, and cannot easily adapt to variations at runtime.


\subsection{ML-based Prefetcher Managers}
ML methods have been gaining traction in place of heuristic methods for achieving superior prefetcher management performance \cite{bhatia2019perceptron, liao2009machine, jimenez2012making, hiebel2019machine, maldikar2014adaptive}. 
ML algorithms can extract the non-intuitive interactions between the different prefetchers. 
Prior methods on prefetcher management configure or train the manager, which typically predicts which PSC to use for a given application, using values of hardware events collected for a single fixed PSC (generally, the default PSC) \cite{liao2009machine, jimenez2012making,  bhatia2019perceptron, rahman2015maximizing}.
Using an ML model trained using only a single fixed PSC would make sense if the prefetcher system always uses that single fixed PSC at runtime.
However, at runtime the PSC changes.
If the values of the hardware events are highly dependent on which PSC is being used, using a dataset generated using a fixed PSC for training leads to a low-accuracy ML model for the prefetcher adaptation.
To address this concern, we train our ML model using PSC-invariant hardware events (i.e., events that are not dependent on the PSC, e.g., the number of branch instructions).

We observe a wide variation in the complexity of the ML algorithms used in prior work. 
We list the prior work in Table \ref{tab:prior}.
Some of the algorithms are simple and either use small datasets or use datasets that do not accurately portray the runtime environment as they use PSC-variant events and only train data of one fixed PSC. 
As a result, these algorithms cannot achieve good accuracy at runtime \cite{liao2009machine, jimenez2012making, rahman2015maximizing}.
Other algorithms, such as neural networks, are too complex and their size increases prohibitively with the size of the dataset \cite{bhatia2019perceptron}.
Moreover, some prior works focus on hardware adaptation only from the perspective of accuracy without worrying about the hardware implementation \cite{liao2009machine, jimenez2012making, hiebel2019machine, rahman2015maximizing}. 

Contrary to the prior work, we jointly account for accuracy of the ML model and hardware overhead when designing \myalgo. 
\myalgo~is complex enough to provide good accuracy on {\color{black}a wide variety of applications.} 
At the same time, \myalgo~is not too complex (as demonstrated in Section \ref{sec:hardware}) to be implemented in hardware and scales well with the size/complexity of the dataset. 
Furthermore, \myalgo~is agnostic of the underlying internal mechanics of the prefetchers and can be easily retrained for a new prefetcher that is introduced in a new system. Additionally, to the best of our knowledge, \myalgo~is the only manager that targets all cache levels in the memory hierarchy.

\subsection{Overview of Existing Prefetchers}
The focus of our work is on a prefetcher manager and our manager can work with any type of prefetcher.
Broadly, we can split prefetching techniques into several categories.
We can have regular-pattern-based, i.e, stride-based prefetchers \cite{kagi1996memory, wenisch2009practical},
irregular-pattern-based, i.e., stream-based prefetchers \cite{cantin2006stealth, somogyi2006spatial},
prefetchers that track both regular and irregular patterns~\cite{shevgoor2015efficiently, kim2016path}, and region-based prefetchers \cite{ganusov2005future,wang2003guided}.
A very good comprehensive survey on prefetchers can be found in Falsafi et al. \cite{falsafi2014primer}.

ML-based algorithms can be used for predicting the addresses of the instructions/data that should be prefetched at each cache level in the memory hierarchy.
The ML-based solutions include table-based reinforcement learning \cite{peled2015semantic, bera2021}, linear model-based reinforcement learning \cite{zhang2018using}, perceptron-based neural networks \cite{peled2019neural, fedchenko2019feedforward}, Markov chain model \cite{moreira2017data} and
LSTM-based neural networks \cite{hashemi2018learning, shi2019neural, shi2021hierarchical, zhang2020raop, narayanan2018deepcache, srivastava2019predicting, braun2019understanding, zeng2017long, rogers2019effects} to predict memory access patterns. 
As ML algorithms get more powerful, and the techniques to compress these algorithms become more sophisticated, ML-based prefetchers will become common place in processors.

\section{\myalgo~Design}
\label{sec:suitap}
\subsection{\myalgo~System-Level Overview}
\label{sec:suitap_overview}

\begin{figure}[t]
  \centering
  \includegraphics[width=0.65\columnwidth]{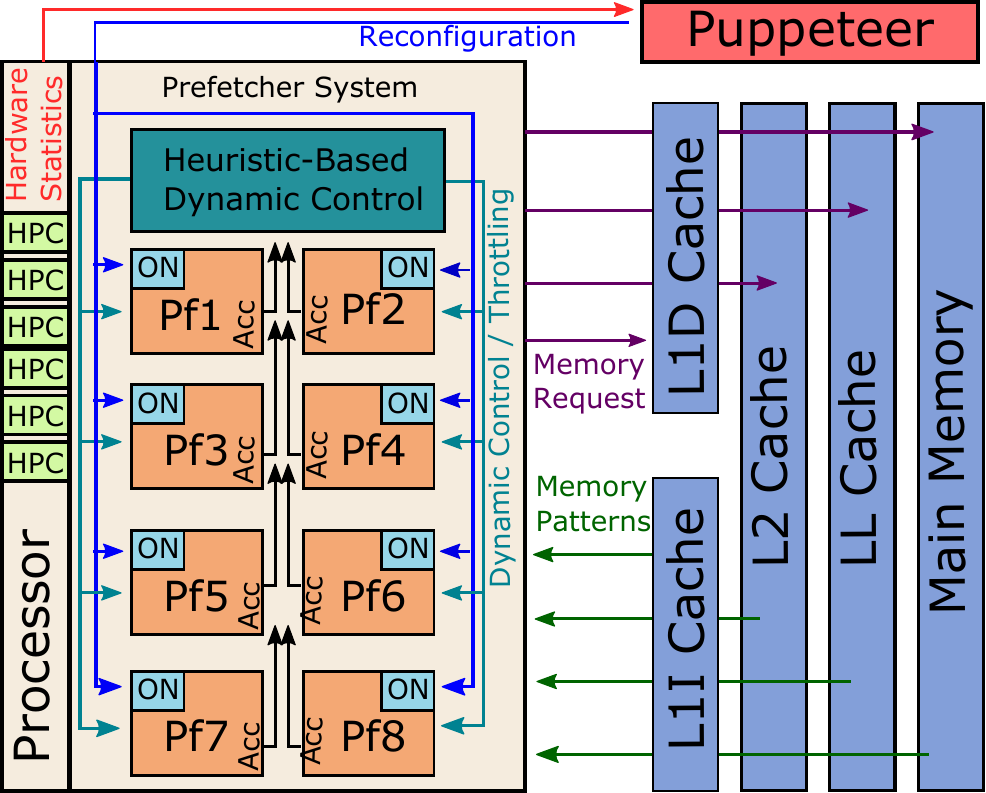}
  \caption{\textbf{Overview of a \myalgo-based System -} Here Pf = prefetcher. Acc = Accuracy of the prefetchers. Pf1 and Pf2 target instructions to bring into L1I\$, Pf3 and Pf4 prefetch data into L1D\$, Pf5 and Pf6 prefetch data into L2\$, while Pf7 and Pf8 target data to prefetch into LLC. Heuristic-Based Dynamic Control block, is a heuristic algorithm that controls the low-level cycle behavior of the prefetchers. \myalgo~controls the longer-term behavior. HPC values are fed into \myalgo~as inputs at runtime.}
  \label{fig:sysarchitecture}
  \vspace{-0.2in}
\end{figure}

In Figure \ref{fig:sysarchitecture}, we show the system-level design of an example prefetcher system that uses \myalgo.
The prefetcher system consists of eight different prefetchers (Pf1 to Pf8), which is typical in modern high performance processors such as Intel i9 \cite{Inteli9} and AMD Ryzen7 \cite{AMDryzen2017}.
These eight prefetchers track different memory access patterns and prefetch data from main memory to last-level cache (LLC), from LLC to L2\$, and from L2\$ to L1\$.
Pf1 and Pf2 target instruction lines to bring into L1I\$, Pf3 and Pf4 prefetch data into L1D\$, Pf5 and Pf6 prefetch data into L2\$, while Pf7 and Pf8 target data to bring into LLC. 
These prefetchers compete among them for cache and memory resources. 
Even across memory levels, a wrong prefetch request from lower levels of memory can harm the performance of prefetchers at higher levels of memory. 
These prefetchers can sometimes act overly aggressive, and can adversely affect each other, in turn leading to loss of application performance. 
There are many heuristics-based algorithms that use simple inputs such as accuracy\footnote{Here accuracy of a prefetcher is quantified as the fraction of the total prefetches that were actually useful, and is calculated as \#prefetches referenced by the program divided by total \#prefetches.
Note, this prefetcher accuracy is not the same as the ML model accuracy. 
Scope is calculated as total \#misses eliminated by the prefetcher divided by the total \#misses when prefetching is disabled.} of the prefetchers or memory bandwidth utilization to throttle prefetchers in such adverse scenarios~\cite{ebrahimi2009coordinated, ebrahimi2009techniques, ebrahimi2010fairness, ebrahimi2011prefetch, heirman2018near, hur2006memory, srinath2007feedback}.
These heuristic algorithms are designed to have low overhead, and hence are highly optimized for the prefetchers in a given prefetcher system.

\myalgo~works as a manager of all prefetchers and complements the heuristic algorithms used in the given prefetcher system.
At runtime, \myalgo~periodically updates the PSC, i.e., it sets which prefetcher should be ON and which should be OFF at each level in the memory hierarchy.
To update the PSC, \myalgo~uses an ML model with the PSC-invariant hardware events, collected from hardware performance counters (HPCs) as inputs.
While low-overhead heuristic algorithms are still required to make extremely low latency decisions at the cycle level, \myalgo~provides additional adaptability by leveraging the power of ML and thereby increasing the performance.
Effectively, {\color{black} heuristic algorithms such as throttling are used in the system to constantly regulate the} short-term behavior of the prefetchers, while \myalgo~controls the longer-term system-level behavior (across hundreds of thousands of cycles). 

\begin{figure*}
  \centering
    \subfloat[Single Classifier]{
    \includegraphics[clip,width=0.3\columnwidth]{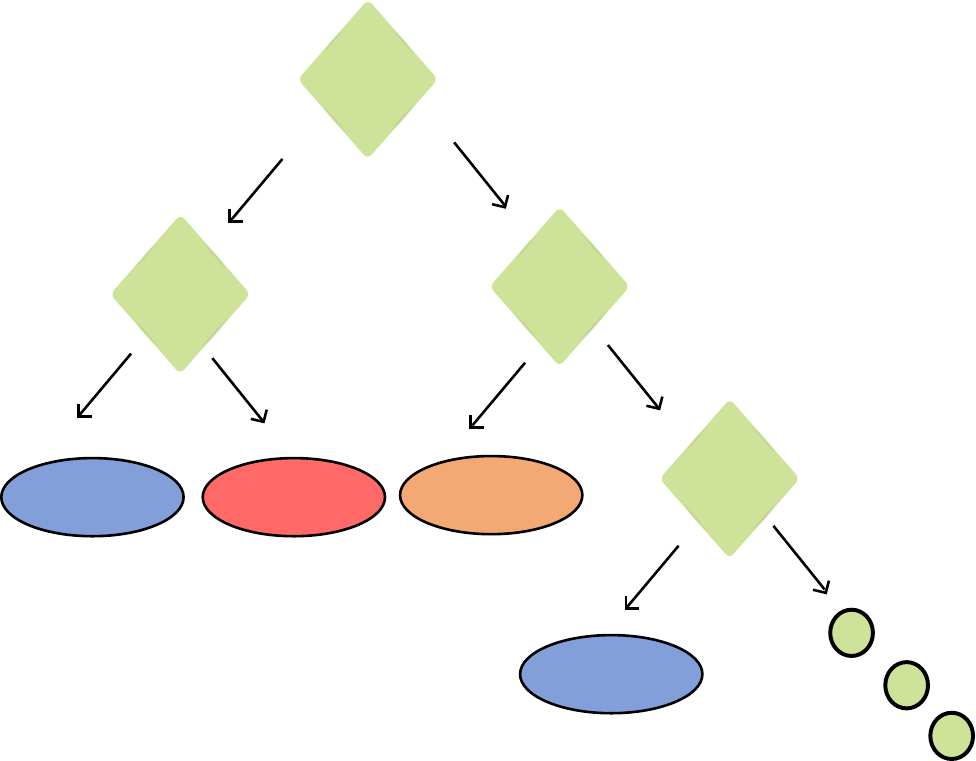}
  }
  \subfloat[Single Regressor]{
    \includegraphics[clip,width=0.3\columnwidth]{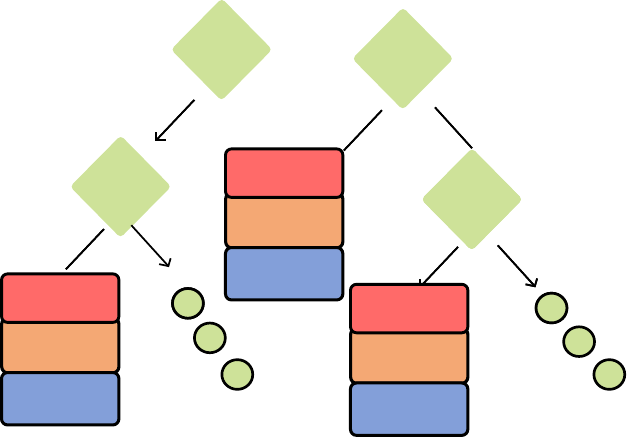}
  }
  \subfloat[Suite of Classifiers]{
    \includegraphics[clip,width=0.6\columnwidth]{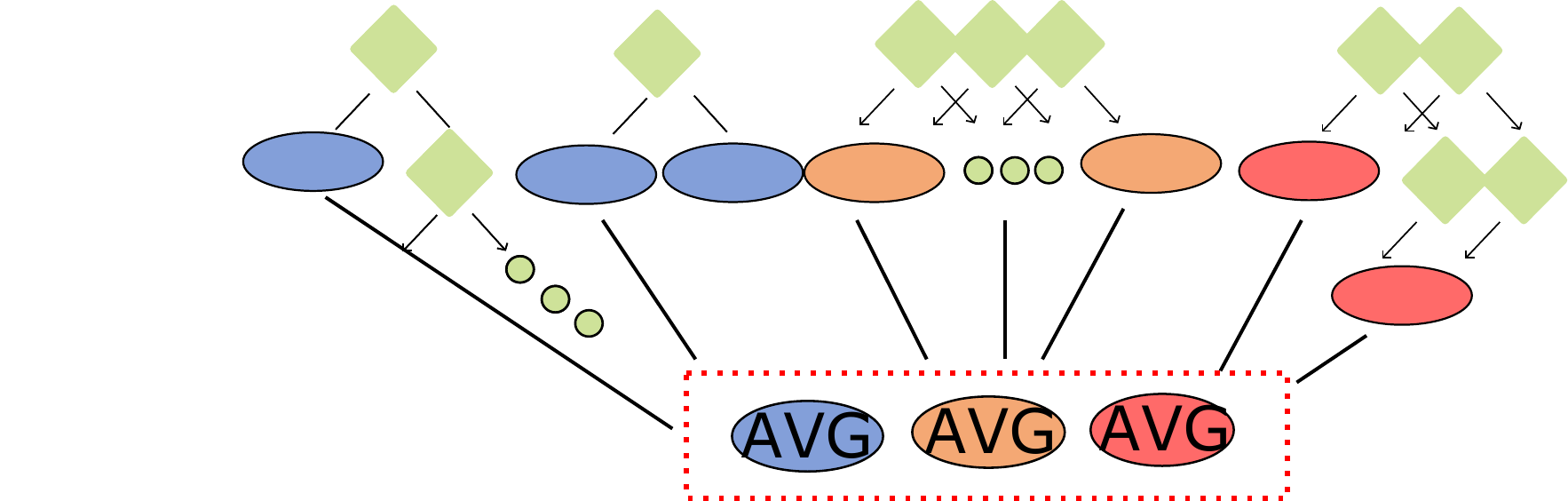}
  }
    \subfloat[Suite of Regressors]{
    \includegraphics[clip,width=0.6\columnwidth]{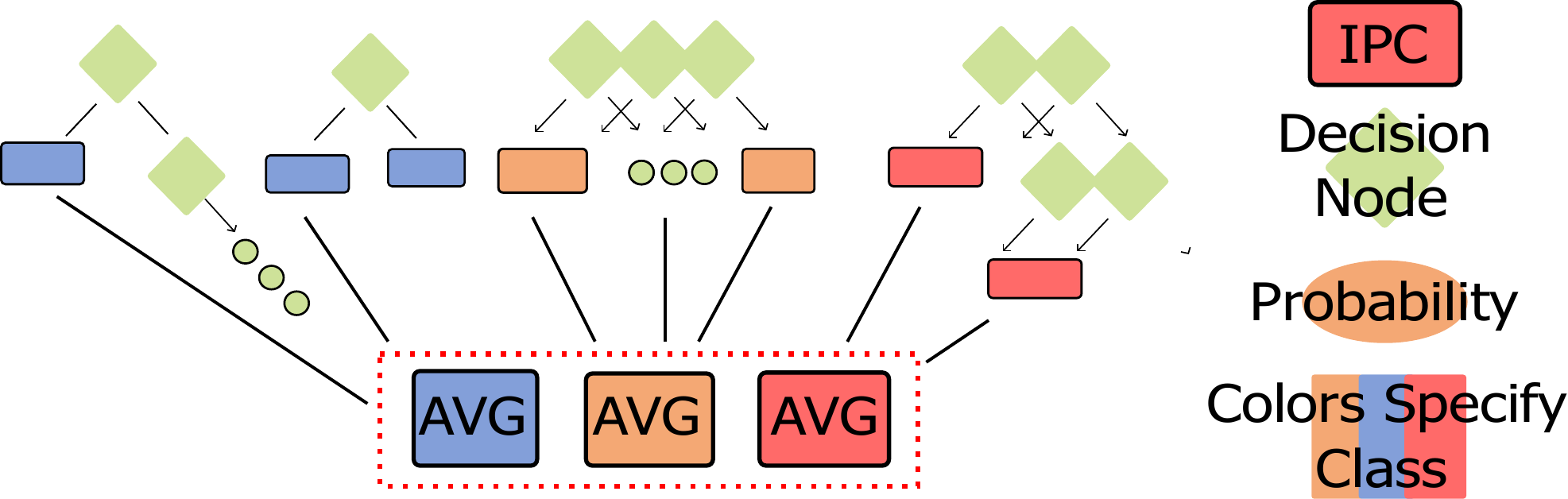}
  }
  \caption{\textbf{Algorithm Options -} A decision tree for a singular classifier, a random forest for singular regressor, a suite of decision trees for a suite of classifiers, and a suite of random forest regressors for a suite of regressors, i.e. \myalgo.}
  \label{fig:method_compare}
  \vspace{-0.2in}
\end{figure*}

\subsection{\myalgo~\suitapmech}
\label{sec:suitap-mech}
For the ML-based \myalgo~algorithm we considered a classification-based approach and a regression-based approach. 
The classification-based approach has been used by prior works because it is relatively easy to train offline and has lower hardware overhead compared to regression.
The regression-based approach has potential for higher performance and has better tolerance to variability compared to classification during runtime when trained offline. 
\ignore{
\subsubsection{Multi-Label vs. Binary Classification}
\label{sec:multi-vs-binary}
Our analysis shows that multiple PSCs are sometimes suitable for a \trace. {\color{black}For example for the \textit{607.cactuBSSN\_s-2421B} trace from SPEC2017, only a few PSCs are suitable. On the other hand, for the \textit{638.imagick\_s-10316B} trace all PSCs are suitable.} Using binary classification, which tries to find the best PSC, can lead to a sub-optimal choice of the PSC. Multi-label classification on the other hand is aware of the multiple suitable PSCs and results in a better choice of PSC. So we use multi-label classification in our ML model.
}

\noindent \textbf{Classification vs. Regression:}
To train \myalgo, as a classification problem we created a dataset using thresholding method similar to prior works~\cite{liao2009machine, jimenez2012making, bhatia2019perceptron, maldikar2014adaptive}.
Here, a trace is run using all available PSCs.
A PSC is given a label of ``1'' if the IPC when using that PSC is within some threshold (in the case of the prior work the threshold is 0.5\%) of the IPC when using the ideal PSC. 
Multiple PSCs can pass the chosen threshold for a given program phase and we then end up using the PSC that is predicted as ``1'' with the highest probability.
Otherwise, the PSC receives a label of ``0''.
Using such a classification approach leads to sub-optimal results.

As an example, consider we have four different traces.
Let us say we classify the first three out of the four traces correctly and the fourth one incorrectly -- i.e. we are able to identify the correct PSC for the first three traces, but not the fourth trace.
So, our classification accuracy is 75\%. 
This means that the first three traces will have performance that is within 0.5\% of their `ideal' performance.
However, the performance of the fourth trace could be 100\% worse or just 0.51\% worse than the `ideal' performance.
This variation in the performance is not accounted in the classification-based model.
Furthermore, this problem is not unique to the given example. 
Any classification-based method would have a similar issue because all labeling methods used to create the dataset for the classification algorithm will certainly lose some amount of information. 

In contrast, a regression-based approach accounts for the \textit{value} of IPC gain/loss and not just if there is IPC gain/loss, when deciding the PSC. 
Given that the regression algorithm is trained on the IPC values directly, the quantitative information of IPC gain/loss is not lost, and the regression algorithm can learn the magnitude of a good or bad prediction.
In particular, the regression algorithm will be used to predict an IPC value for each PSC for a given instruction window. Then, we choose the PSC with the highest predicted IPC value.

\noindent \textbf{Suite of Regressors vs Single Regressor:}
When using regression algorithms, we have two options: (i) use a single regressor, where all data collected from all the PSCs are used to train that single regressor; or (ii) use a suite of regressors, where each PSC will have a dedicated regressor.
Using a suite of regressors leads to a more customized solution that has higher accuracy as compared to using a single regressor. 
Conceptually, this is because in a single regressor, we maximize the accuracy across all PSCs instead of maximizing the accuracy of each PSC separately, whereas, in a suite of regressors we train a dedicated regressor for each PSC. 
In this way, we indirectly jointly increase the scope and accuracy of the overall prefetching system.

In our work, we use a suite of regressors where we implement each regressor using random forest, i.e. one random forest trained per PSC, due to its simple implementation, its robustness to noise in the dataset, its lower overhead (compared to other ML algorithms such as neural networks), and its higher accuracy (compared to other ML algorithms such as decision trees) \cite{kursa2014robustness,
oshiro2012many}. 

We have multiple trees per forest and we allow each tree to split at locations that are unique to the PSC associated with the forest.
The leaves of each tree in the forest specify the predicted IPC value for the PSC.
For each forest i.e. each PSC, we calculate the average of the predicted IPC values obtained from all the trees in the forest, and then choose the PSC with the highest average predicted IPC.
Given that each forest has multiple decision trees, our method has higher tolerance to wrong decisions by the trees, where even if some of the trees give wrong decisions, other trees can compensate.

In Figure \ref{fig:method_compare}, we conceptually show the differences between the four different options discussed above: (a) single classifier, (b) single regressor (c) a suite of classifiers, and (d) a suite of regressors. Respectively, the leaf nodes in (a) contain the PSC choice directly, (b) have the predicted highest IPC among all the PSCs, (c) the probability value of a given instruction window belonging to the given PSC (of which we choose the highest one), and (d) the predicted IPC value for a given PSC which will be averaged per tree in a given RF and compared with the other averaged predicted PSC values from the other RFs (of which we will choose the highest). For (a) and (b) the PSCs are traversed simultaneously, since the PSCs share a single algorithm, while for (c) and (d) each PSC has a unique algorithm we traverse.

\subsection{\myalgo Training}
\label{sec:training_details}
To train \myalgo we need to generate a representative dataset.
Consider the case where we have a single prefetcher, $Pf$, {\color{black}at only one level in the memory hierarchy}.
Here the number of PSCs ($N_{psc}) = 2$, i.e., $Pf$=OFF and $Pf$=ON.
For two consecutive instruction windows, we will have $N_{psc}^2 = 4$ possible scenarios: (i) $Pf$=OFF $\rightarrow$ $Pf$=OFF, (ii) $Pf$=ON $\rightarrow$ $Pf$=OFF, (iii) $Pf$=OFF $\rightarrow$ $Pf$=ON, and (iv) $Pf$=ON $\rightarrow$ $Pf$=ON.
With $N$ number of instruction windows and $N_{trace}$ number of traces, the number of different possible scenarios will then be $N_{trace} \times N_{psc}^N$.
When $N$ increases, the number of different scenarios will increase exponentially, hence including each unique scenario in the dataset for training is not feasible.
To handle this problem, we propose to use only \mbox{PSC-invariant} events as our features. 
An example of a PSC-invariant event is the number of conditional branches, which is not affected by the choice of PSC. 
We check the variance of each hardware event value (for 180 total hardware events that we can track) for each PSC. 
We identify 59 events whose values vary by less than $\pm$10\% from their mean value across all PSCs. 
We further reduce the number of events by eliminating the redundant events that track similar behavior and have high correlation with each other.
Table \ref{tab:arch} shows the final 6 events we choose to track trace behavior.
After we have identified our PSC-invariant events that will be the features and the PSCs that will be the choices of our ML model, we collect an IPC value per PSC for each instruction window as our ground truth. 

\ignore{
After we have identified our PSC-invariant events that will be the features and the PSCs that will be the classes of our ML model, we run the $232$ \trace s for 1.2B instructions on 1 core and collect the PSC-invariant event values using 100k instruction windows.
We use a warm up phase of 200M instructions before we start collecting the PSC-invariant values. 
In total we have 232 (traces)$\times$10,000 (instruction windows per trace) = 2,320,000 instructions windows that make up our dataset.
In the dataset we have an IPC value per PSC for each instruction window as our labels.}

Using the features and IPC values we have collected, we then form our suite of random forest regressors wherein we train a separate forest for each PSC using CART (classification and regression trees) \cite{steinberg2009cart}. 
CART is a greedy recursive search algorithm that maximizes information by splitting the data at each node using one feature. Each child node is split recursively until there is no information gain from splitting a child node.
We limit the total number of decision nodes in \myalgo~to keep the size of \myalgo~smaller than L1\$. 
With this limitation in mind, we conduct a hyper-parameter search and determine that the number of estimators (trees per random forest) should be 5 and the number of max nodes should be 100 per tree.

\subsection{\myalgo~Microarchitecture Design}
\label{sec:hardware}

\begin{figure}[t]
  \centering
  \includegraphics[width=0.7\columnwidth]{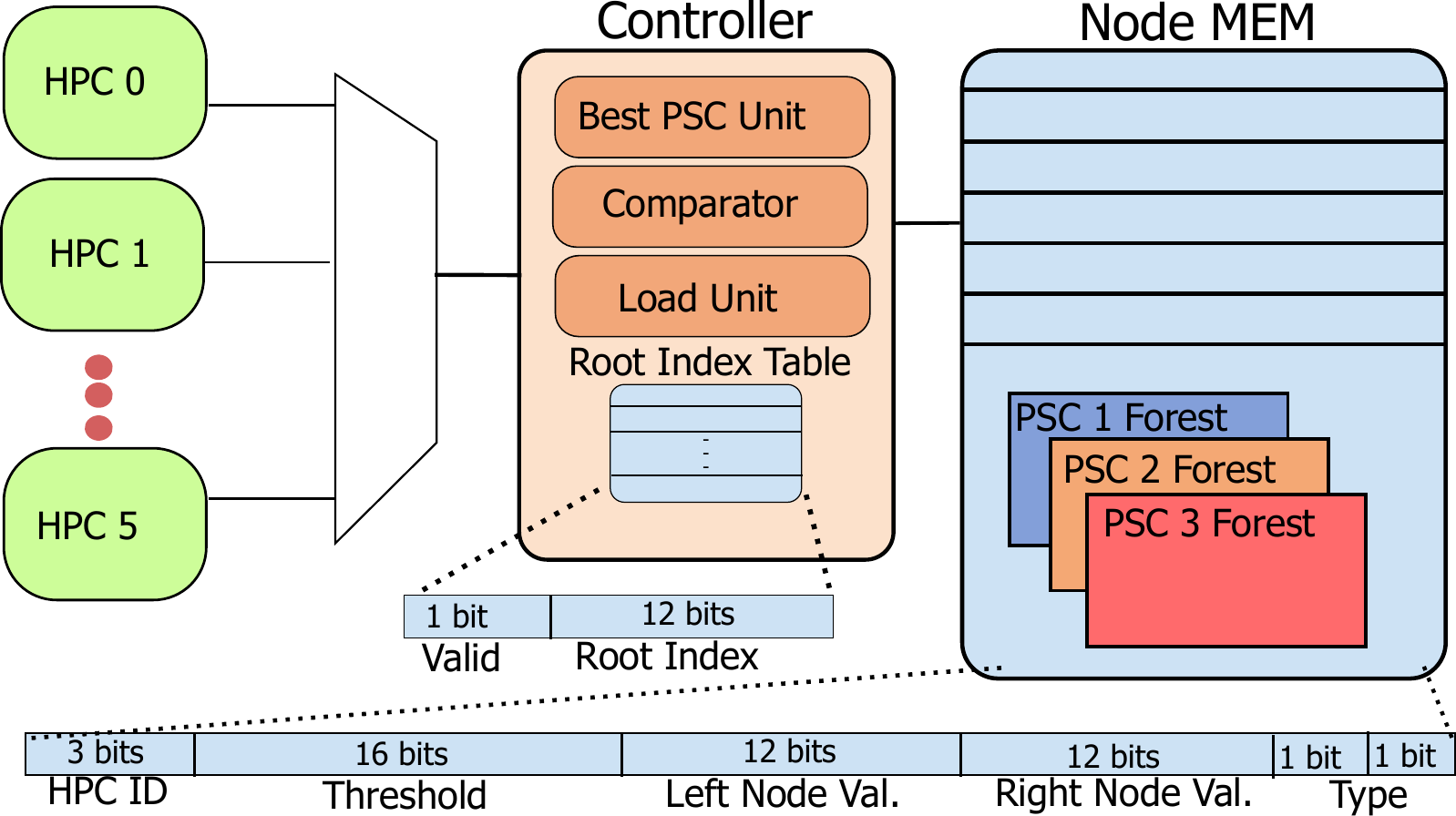}
  \caption{\textbf{\myalgo~Hardware Design -} \myalgo is made up of a Node MEM - SRAM array, a max logic unit, and several register files.}
  \label{fig:structure}
  \vspace{-0.25in}
\end{figure}
\ignore{
We use a total of 2250 nodes, which allows us to store \myalgo~in an SRAM array of a smaller size than L1\$, to design the suite of random forest regressors. We use a single port SRAM array called \textit{Node MEM} to store information about {\color{black}all the 2250 nodes.}} 

Figure \ref{fig:structure} shows the microarchitecture details of \myalgo.
We use a single port SRAM array called \textit{Node MEM} to store information about the nodes that form the trees of each random forest in \myalgo.
We load the random forest-based Puppeteer model into the \textit{Node MEM} at startup using firmware.
Each entry of \textit{Node MEM} corresponds to one node in one of the random forests and it consists of the following fields:
(i) A 3-bit \texttt{HPC ID} field that specifies which PSC-invariant event, i.e. which hardware performance counter (HPC), is used by that node to make a decision. 
The 3-bit encoding enables the node to use one of 6 different PSC-invariant events (see Table~\ref{tab:arch}).
(ii) A 16-bit \texttt{Threshold} field (threshold value is determined during training), {\color{black}which is employed {\color{black}by the node} to decide if the decision path should branch left or right}.
In our problem 16 bits provide enough precision for the ML model weight values.
(iii) A 12-bit (for 2250 node addresses) \texttt{Left~Node~Value(LNV)} field, and (iv) a 12-bit \texttt{Right~Node~Value(RNV)} field.
These \texttt{LNV} and \texttt{RNV} fields represent child node indices for internal nodes of a tree.
For the leaf nodes of a tree, we use these \texttt{LNV} and \texttt{RNV} fields to indicate the predicted IPC value of a PSC.
We differentiate between child node index and predicted IPC using (v) a 1-bit \texttt{Type} field. 
We use a separate 1-bit \texttt{Type} field for \texttt{LNV} and \texttt{RNV}.

At the end of every instruction window, \myalgo~calculates the predicted IPC for each PSC in the next instruction window by traversing the trees of the associated forest and using the PSC-invariant event values for the current window as inputs. 
For each forest, the controller in \myalgo~reads the \textit{Node MEM} index of the root {\color{black}node} for the first tree from \textit{Root Index Table (RIT)} and loads the \textit{Node MEM} entry for {\color{black}the} root node using a \textit{Load Unit} into a register. 
Next, the \texttt{HPC} \texttt{ID} in the loaded \textit{Node MEM} entry is used to load the corresponding PSC-invariant event value into a second register. 
Then the \texttt{Threshold} value, stored in the first register, and PSC-invariant event value stored in the second register are compared using the \textit{Comparator}.
Based on the \textit{Comparator} output, we choose to traverse down to the left child or the right child.
The \textit{Controller} then uses the {\color{black}corresponding index value from \texttt{LNV} or \texttt{RNV}} to find the next node in \textit{Node MEM}.
The \textit{Controller} continues traversing the tree until it loads a predicted IPC value corresponding to a leaf from the \textit{Node MEM}. 
The above steps are repeated for the remaining trees in the forest, and then we calculate the average of the predicted IPC values obtained from all the trees in that forest.
The \textit{Best PSC Unit} in the \textit{Controller} stores the ID of the PSC with the highest predicted IPC value.
Every time the \textit{Controller} finishes traversing a forest, the predicted IPC value of that forest, i.e. PSC, is compared with the predicted IPC value stored in the \textit{Best PSC Unit} using the \textit{Comparator}.
If the new predicted IPC value is higher than the current value, the \textit{Best PSC Unit} updates the predicted IPC value and the ID of the PSC.
Once all forests have been traversed i.e., all PSCs have been evaluated, \myalgo~chooses the entry stored in the \textit{Best PSC Unit} as the PSC for the next instruction window.

\ignore{
At the end of an instruction window we do not have to stall the processor until \myalgo~figures out the PSC for the next window.
\myalgo~can determine the best PSC while
instructions are being executed, and we can update the PSC in the background without impacting the on-going execution. This
is possible because when changing the PSC we simply
allow or disallow (gate) outgoing prefetch requests based on whether the
prefetcher is ON or OFF. We do not enable or disable the prefetchers. This
choice also has an added benefit of allowing gated prefetchers to still get
partially trained at runtime and to stay prefetch ready.}

We determined that a maximum depth of $10$ per tree is more than sufficient to accurately determine the best PSC. 
In our evaluation we use a prefetcher system with $N_{psc}$=5 (given in Table \ref{tab:final_psc} and discussed in detail in Section~\ref{sec:methodology}). We need a total of 2250 nodes to design the trees in \myalgo, and these nodes require a {\color{black}10.75} KB-sized \textit{Node MEM} (compared to a typical L1\$ of 32 KB).
Other than \textit{Node MEM}, we require a 5$\times$N$_{psc}$-entry \textit{RIT} where each entry is 13-bit wide (12 bits for the root node index and 1 valid bit), a 12-bit comparator containing comparison logic and two registers, a load unit, and a register to store the best PSC information in the \textit{Controller}. We discuss the hardware overhead in more detail in Section \ref{sec:results}.

\section{Evaluation Methodology}
\label{sec:methodology}

\begin{table}[t]
  \centering
  \vspace{0.05in}
  \caption{\textbf{Simulated System Parameters}}
  \vspace{-0.15in}
  \label{tab:arch}
\scriptsize
\renewcommand{\arraystretch}{1.0}
\centering
\center{
\begin{tabular}{l|l|l|l|l}
\hline
\hline
\multicolumn{1}{c|}{\textbf{Component}} &  \multicolumn{4}{P{6.5cm}}{\textbf{Simulated Parameters}}\\
\hline \hline
\multicolumn{1}{c|}{Core} & \multicolumn{4}{P{6.5cm}}{One to four cores, 4GHz, 4-wide, 256-entry ROB}\\
\hline
\multicolumn{1}{c|}{TLBs} & \multicolumn{4}{P{6.5cm}}{64 entries ITLB, 64 entries DTLB, 1536 entry shared L2 TLB} \\
\hline
\multicolumn{1}{c|}{L1I\$} & \multicolumn{4}{P{6.5cm}}{32KB, 8-way, 3 cycles, PQ: 8, MSHR: 8, 4 ports} \\
\hline
\multicolumn{1}{c|}{L1D\$} &  \multicolumn{4}{P{6.5cm}}{48KB, 12-way, 5 cycles, PQ: 8, MSHR: 16, 2 ports} \\
\hline
\multicolumn{1}{c|}{L2\$} & \multicolumn{4}{P{6.5cm}}{512KB, 8-way, 10 cycles, PQ: 16, MSHR: 32, 2 ports} \\
\hline
\multicolumn{1}{c|}{LLC} & \multicolumn{4}{P{6.5cm}}{2MB/core, 16-way, 20 cycles, PQ: 32×cores, MSHR: 64×cores} \\
\hline
\multicolumn{1}{c|}{DRAM} & \multicolumn{4}{P{6.5cm}}{4GB 1 channel/1-core, 8GB 2 channels/multi-core, 1600 MT/sec}
\\
\hline \hline \hline
\multicolumn{2}{l|}{\textbf{Hardware Event}} &  \multicolumn{3}{l}{\textbf{Properties}}\\
\hline \hline
\multicolumn{2}{l|}{$L1I\_PAGES\_READ\_LOAD$} & \multicolumn{3}{l}{L1I\$ Pages Read on Load.}\\
\hline
\multicolumn{2}{l|}{$L1D\_PAGES\_READ\_LOAD$} & \multicolumn{3}{l}{L1D\$ Pages Read on Load.}\\
\hline
\multicolumn{2}{l|}{$L1D\_RFO\_ACCESS$} & \multicolumn{3}{l}{L1D\$ Store Accesses.} \\
\hline
\multicolumn{2}{l|}{$BRANCH\_RETURN$} & \multicolumn{3}{l}{Branch Returns.} \\
\hline
\multicolumn{2}{l|}{$NOT\_BRANCH$} &  \multicolumn{3}{l}{Not Branches.} \\
\hline
\multicolumn{2}{l|}{$BRANCH\_CONDITIONAL$} & \multicolumn{3}{l}{Conditional branches.}
\\
\hline \hline 
\end{tabular}
}
\end{table}

\begin{table}[t]
  \centering
  \caption{\textbf{Static PSCs and Managers Evaluated -} We list the static PSCs from the prior prefetching competitions, the manager algorithms from prior work, and the different flavors of \myalgo.}
  \vspace{-0.1in}
  \label{tab:algorithms}
\scriptsize
\centering
\center{
\begin{tabular}{P{0.4in}|p{1.8in}|P{0.5in}|P{0.3in}}
\hline
\hline
\textbf{Algorithm Notation} & \textbf{Explanation} & \textbf{Static PSC or Manager} & \textbf{Training Dataset} \\
\hline
NO & No prefetching is used. PSC = \textit{no-no-no-no} & Static PSC & Not trained \\
\hline
IPCP & Winner of DPC3 \cite{dpc3}. PSC = \textit{no-ipcp-ipcp-nl}. & Static PSC & Not trained \\
\hline
EIP & Winner of IPC1 \cite{ipc1}. PSC = \textit{EIP-nl-spp-no}. & Static PSC & Not trained \\
\hline
PY & RL-based algorithm named Pythia \cite{bera2021} & ML-based Prefetcher & Online \\
\hline
J3 & Heuristic-based algorithm  \cite{jimenez2012making} & Manager & Not trained \\
\hline
NN & Multi-layer-perceptron  \cite{bhatia2019perceptron} & Manager & 1C \\
\hline
B1C & Decision tree algorithm by Liao et al. \cite{liao2009machine} & Manager & 1C \\
\hline
B4CS & Decision tree algorithm by Liao et al. \cite{liao2009machine} & Manager & 4CS \\
\hline
B4CM & Decision tree algorithm by Liao et al. \cite{liao2009machine} & Manager & 4CM \\
\hline
P1C & \myalgo  & Manager & 1C \\
\hline
P4CS & \myalgo & Manager & 4CS \\
\hline
P4CM & \myalgo & Manager & 4CM \\
\hline
\hline
\end{tabular}
}
\vspace{-0.2in} 
\end{table}

{\color{black}
We use ChampSim \cite{champsim} for our analysis, where we model one core (1C), four core (4C), and eight core (8C) processors to have multiple prefetchers at each level of the cache -- private L1I\$, L1D\$, private L2\$ and shared LLC (more details provided in Table~\ref{tab:arch}).
We train \myalgo using data collected from a 1C and 4C OoO processor and then evaluate it on 1C, 4C, and 8C OoO processors. 
In the 4C and 8C processors, we have a private \myalgo~per core. 
Each private \myalgo~utilizes the 6 PSC-invariant events per core and makes independent decisions. 
} 
For our evaluation we use a diverse set of 232 traces generated from SPEC2017 \cite{SPEC2017}, SPEC2006 \cite{SPEC2006}, and Cloud \cite{ipc1} benchmarks. 

In Table~\ref{tab:algorithms} we list the notations used for all the static PSCs and the runtime managers (that change PSC at runtime) that we have evaluated.
In our evaluation, we normalize all IPC values to the same state-of-the-art baseline as PPF~\cite{bhatia2019perceptron}, i.e., SPP~\cite{kim2016path} (\textit{no-no-spp-no})\ignore{\footnote{PSC=4-tuple: $L1I\$_{prefetcher}-L1D\$_{prefetcher}-L2\$_{prefetcher}-LLC_{prefetcher}$.})}. 
We compare \myalgo~against the best static PSCs from competitions, i.e., IPCP (\textit{no-ipcp-ipcp-nl}) and EIP (\textit{EIP-nl-spp-no}; as well as managers such as the final version of the algorithm developed by Jimenez et al. that uses trial periods to latch onto the PSC \cite{jimenez2012making} (J3), Pythia that is an RL-based algorithm developed by Bera et al. \cite{bera2021} (PY)\footnote{Note that Bera et al. used SHiP \cite{wu2011ship} for their cache replacement policy and perceptron as their branch predictor. We tested with their settings as well as with using Pythia with hashed-perceptron and LRU, and observed on average that Pythia with hashed-perceptron with LRU has 12\% higher IPC. Hence, we use hashed-perceptron and LRU similar to all the other comparison points in this paper.}, a multi-layer-perceptron similar to Bhatia et al. \cite{bhatia2019perceptron} (NN), and a binary-tree (BT) based algorithm \cite{liao2009machine} - where Liao et al. tried several different ML methods (such as decision trees and NN) and concluded that decision trees are the best choice. 

\subsection{Training Approaches for ML-based Prefetcher Managers}
When evaluating any new idea, a widely used approach in the industry is to use all available evaluation data.
We are using a ML-based approach and are cognizant of the fact that we do not want to overfit our model using all available data for training the ML model.
So here, we compare three different approaches for constructing the training dataset for \myalgo.
For training, we have 232 traces$\times$10,000 instruction windows per trace = 2,320,000 instructions windows.
For each approach, we limit the training data in different ways. 
Note that we are still training at the instruction window-level and performing 10-fold cross-validation on the training set for all approaches.
We give the percentage of data used to construct each training dataset in Table \ref{tab:dataset_approaches}.

In the \textbf{first approach}, we severely limit the data and randomly select 10\% of the instruction windows at random time intervals to form the training dataset, i.e., 90\% of the instruction windows are not seen during training.
The idea behind this approach is that we are training the algorithm to recognize low-level memory access behavior.
This approach does not overfit the model because the behavior of the benchmarks when collecting the training data would be different from the behavior of the benchmarks when we use the trained \myalgo.
This difference is because when collecting training data we do not change the PSC, while when using \myalgo the PSC can potentially change for each instruction window.
In the \textbf{second approach}, we generate the dataset by leaving 20\% of the traces out of the training dataset and using 60\% of the instruction windows from the remaining 80\% traces for training.
We use a 60-40 split of the instruction windows to avoid overfitting the model to the 80\% of traces in the training set.
Given traces are constructed to represent unique behaviors in each application, our training set will not have information on all the unique behaviors of a given benchmark.
This means that some of the traces from other benchmarks are assumed to be from the same distribution of these missing traces or else there will be no way for the algorithm to train for these regions. 
In the \textbf{third approach}, we generate the dataset by leaving 20\% of the benchmarks out of the training dataset and using a 60-40 split of the instruction windows belonging to the remaining 80\% benchmarks for training.
In this approach, whole benchmarks are not considered during training.
Here too, we choose a 60-40 split of the instruction windows to avoid overfitting the model to the 80\% of benchmarks in the training set.
Similar to the second approach it is assumed that some of the other benchmarks will be similar in behavior to the missing benchmarks and cover their memory access behavior.  

\ignore{In Figure \ref{fig:1C_unseen}, we show the performance distribution of P1C, B1C, and NN where the training dataset for the managers were generated using the three different approaches. 
Here we show the average IPC of P1C, B1C and NN models on the benchmarks that were not available during training.
For B1C\_IW, B1C\_T, and B1C\_B we observe comparable average IPC gain of 8.2\%, 8.3\%, and 8.8\%, respectively, over SPP. 
We also observe that for all three approaches 30\% of the traces have performance loss compared to SPP with the worse case loss at 26\%, 26.5\%, and 28\%, respectively. 
For NN\_IW, NN\_T, and NN\_B we observe an average IPC gain of 4.2\%, 6.8\%, and 6.8\%, respectively, over SPP.
In addition, ~40\%, 35\% and 35\% of the traces observe performance loss for NN\_IW, NN\_T and NN\_B compared to SPP, and the corresponding worst case loss is 12\%, 10\%, and 10.5\%, respectively.
Effectively, in both B1C and NN, the first training approach has lower IPC gain than the other two approaches.
\myalgo\_IW has 3.1\% and 7.1\% average IPC gain over B1C\_IW and NN\_IW. 
}

In a 1C processor, when using \myalgo~on unseen benchmarks, for the first, second, and third approaches, we observe an average IPC gain of 11.3\%, 9.8\%, and 6.6\%, respectively, over SPP.
In the second and third approaches, we see that \myalgo has lower performance gain than the first approach.
Despite having 3\% more instruction windows for training in the third approach compared to the second approach there is a 3.2\% drop in performance.
Furthermore, in the first approach we use only 10\% of the total number of instruction windows for training (5 $\times$ less instruction windows than the other approaches), yet this approach performs 1.5\% and 4.7\% better than the other two approaches.
This implies that the second and third datasets is trained on an insufficient number of unique memory accesses patterns, leading to low performance observed during runtime execution. 
Therefore for the rest of our evaluation we train all ML-based prefetcher managers using the first approach. 
As mentioned earlier, \myalgo can always be retrained and updated via firmware.


\subsection{1C, 4C and 8C Workload Formulation}

We run 1C and 4C experiments using a 200M instruction warmup phase and 1B instruction detailed simulation phase, while for 8C we use 200M instruction warmup phase and 250M instruction detailed simulation phase.
We generate a total of 232 traces from SPEC2017, SPEC2006, and Cloud benchmarks.
We create two flavors of trace sets for the 4C and 8C experiments:  
a single-type trace set where each core runs the same trace, and a mixed-type trace set where each core runs a unique trace. 
Therefore we have 5 data suites in total: 1C, 4C single-type trace set (4C STS), 4C mixed-type trace set (4C MTS), 8C single-type trace set (8C STS), and 8C mixed-type trace set (8C MTS).
We use hashed-perceptron for branch predictor and least-recently used (LRU) policy for cache replacement policy provided by ChampSim. 
To construct our mixed-type trace sets, we first split the 232 traces into 6 groups based on ascending execution latency.
Then we randomly select a trace from a given group per core and construct a mixed-type trace group.
We cover all permutations of the various latency groups.
This way we have diversity in the traces running on the cores.
For example, in a 4C processor, we can assign a trace from group 2 to core0, a trace from group 3 to core1, a trace from group 4 to core3, and a trace from group 5 to core3.
This can be represented as $2-3-4-5$.
Therefore we have 360 experiments (6 options for core0 $\times$ 5 options for core1 $\times$ 4 options for core2 $\times$ 3 options for core3) in 4C MTS for 6 latency groups and 4 selected traces (1 for each core). 
For 8C MTS, since the number of experiments increases and the experiments take quite long, we replicate the same latency group for 4 of the cores. For example, we can use $1-1-1-1-4-4-4-4$ for an 8C experiment. 
However, instead of 6 groups as in the 4C system, we use 10 groups for the 8C system, to cover a more granular variety of behavior and instead of permutations we use combinations with replacement. 
Therefore, we have 55 (2 traces selected from 10 latency groups with replacement, i.e., $C^R(10,2)$) experiments for 10 groups and 2 selections. 
Note that, even if the latency group is the same, since we choose a trace randomly from a latency group, the trace can still be different for each core.

\begin{table}[]
  \centering
  \vspace{0.08in}
  \caption{\textbf{Dataset training approaches.} Here \% Benchmarks indicates the percentage of benchmarks were used during training, \% Traces indicates the percentage of traces from the available benchmarks were used during training, and \% Inst. Win. indicates the percentage of instructions windows from the the available traces were used during training. The last columns indicates overall the percentage of instruction windows were using during training.}
  \vspace{-0.1in}
  \label{tab:dataset_approaches}
\scriptsize
\renewcommand{\arraystretch}{1.0}
\centering
\center{
\begin{tabular}{P{0.5in}|P{0.6in}|P{0.4in}|P{0.3in}|P{0.9in}}
\hline
\hline
\textbf{Approach} & \textbf{\%Benchmarks} & \textbf{\%Traces} & \textbf{\%Inst. Win.} & \textbf{Total \% of all Inst. Win. used for training} \\
\hline
\# 1 & ALL & ALL & 10\% & 10\%\\
\hline
\# 2 & ALL & 80\% & 60\% & 48\% \\
\hline
\# 3 & 80\% & ALL & 60\% & 51\% \\
\hline
\hline
\end{tabular}
}
\vspace{-0.1in}
\end{table}

\begin{table}[t]
  \centering
  \vspace{0.1in}
  \caption{\textbf{Initial Prefetcher Options -} We show the regular and irregular prefetcher options we used at each cache level to construct our set of 300 PSCs. We reduce the number of PSCs down to 5 PSCs before training.}
  \vspace{-0.1in}
  \label{tab:evaluated-solutions}
\scriptsize
\renewcommand{\arraystretch}{1.0}
\centering
\center{
\begin{tabular}{l|l|l|l}
\hline
\hline
\textbf{L1I\$} & \textbf{L1D\$} & \textbf{L2\$} & \textbf{LLC} \\
\hline
No Prefetcher & No Prefetcher & No Prefetcher & No Prefetcher \\
\hline
Next-line \cite{falsafi2014primer} & Next-line & Next-line & Next-line \\
\hline
FNL+MMA \cite{seznec2020fnl+} & IPCP \cite{Pakalapati2020ipcp} & IPCP & \\
\hline
DJOLT \cite{nakamurad} & MLOP \cite{shakerinava2019multi} & SPP \cite{kim2016path} & \\
\hline
EIP \cite{ros2020entangling} & Bingo \cite{bakhshalipour2019bingo} & KPCP \cite{kim2017kill} & \\
\hline
 &  & Ip-Stride \cite{falsafi2014primer} & \\
\hline \hline 
\end{tabular}
}
\end{table}

\subsection{Generating ML Models and PSC Pruning}
To generate the 1C dataset (PSC and associated IPC values), we run the 1C experiments using all possible PSCs generated from the prefetcher options that are available in the ChampSim repository, the 1st place (IPCP \cite{Pakalapati2020ipcp}), 2nd place (Bingo \cite{bakhshalipour2019bingo}), and 3rd place (MLOP \cite{shakerinava2019multi}) winners of the 3rd data prefetching competition (DPC3) \cite{dpc3}, and the 1st place (EIP \cite{ros2020entangling}), 2nd place (FNL+MMA \cite{seznec2020fnl+}), and 3rd place (DJOLT \cite{nakamurad}) winners of the 1st instruction prefetching competition (IPC1) \cite{ipc1}. 
To reduce the hardware overhead of \myalgo, we avoid including multiple PSCs that cover the same traces.
To this end, we initially ran 20 traces for 20M instructions with all possible PSCs (5 prefetching options in L1I\$ $\times$ 5 prefetching options in L1D\$ $\times$ 6 prefetching options in L2\$ $\times$ 2 prefetching options in LLC = 300 PSCs). 
We summarize the different prefetchers that we evaluated in Table~\ref{tab:evaluated-solutions}.

\begin{table}[t]
  \centering
  \caption{\textbf{PSCs and their prefetching options used at each \$ level.} Note, these PSCs are used by all the manager algorithms not just \myalgo.}
  \vspace{-0.1in}
  \label{tab:final_psc}
\scriptsize
\renewcommand{\arraystretch}{1.0}
\centering
\center{
\begin{tabular}{l|l|l|l|l}
\hline
\hline
\textbf{Final PSCs Used} & \textbf{L1I\$} & \textbf{L1D\$} & \textbf{L2\$}     & \textbf{LLC} \\
\hline
djolt-bingo-nl-nl & DJOLT \cite{nakamurad} & Bingo \cite{bakhshalipour2019bingo} & Next-line \cite{falsafi2014primer} & Next-line\\
\hline
djolt-bingo-no-no & DJOLT & Bingo & No Prefetcher & No Prefetcher\\
\hline
fnl-bingo-spp-nl & FNL+MMA \cite{seznec2020fnl+} & Bingo & SPP \cite{kim2016path} & Next-line\\
\hline
fnl-bingo-spp-no & FNL+MMA & Bingo & SPP & No Prefetcher\\
\hline
no-nl-spp-no & No Prefetcher & Next-line & SPP & No Prefetcher\\
\hline
\textbf{Overhead} & 221KB & 48.06KB & 6KB & 0.6KB \\
\hline \hline
\end{tabular}
}
\vspace{-0.1in}
\end{table}

For each trace, we sort the PSCs based on the corresponding IPC values in descending order. 
We generate a new table for each trace, where the table contains the top 10 PSC entries for the trace and combine these tables to form a super-table that contains the top 10 PSCs for all traces.
Note that a PSC may be in the top 10 for more than one trace.
We sort the PSCs in descending order based on the number of traces for which the PSC is in the top 10.
Starting from the top, we select just enough PSCs to improve performance of all the 20 traces.
We picked the PSCs that have the best performance with minimal coverage overlap while reducing the number of unique prefetchers (to reduce the hardware overhead).
In Table \ref{tab:final_psc} we show the final 5 PSCs that we selected.
These PSCs give good performance for the maximum number of traces. 
We show the prefetcher used at each cache level.
It is interesting to note that the best Prefetchers from DPC3 and IPC1 -- IPCP and EIP -- are not used in the top 5 choices for PSCs. 
This shows that the state-of-the-art prefetchers designed in isolation may not be the best choice of prefetchers when used with other prefetchers.
We collect the 4C STS and 4C MTS datasets, in the same way as the 1C dataset -- the only difference is that the data is collected per core.
We then train BT and \myalgo~using datasets collected from 1C, 4C STS, and 4C MTS. 
We denote the different flavors of the two algorithms as P1C, P4CS, and P4CM for \myalgo; and B1C, B4CS, and B4CM for BT. 

\ignore{
\begin{figure}[t]
  \centering
  \includegraphics[width=0.9\columnwidth]{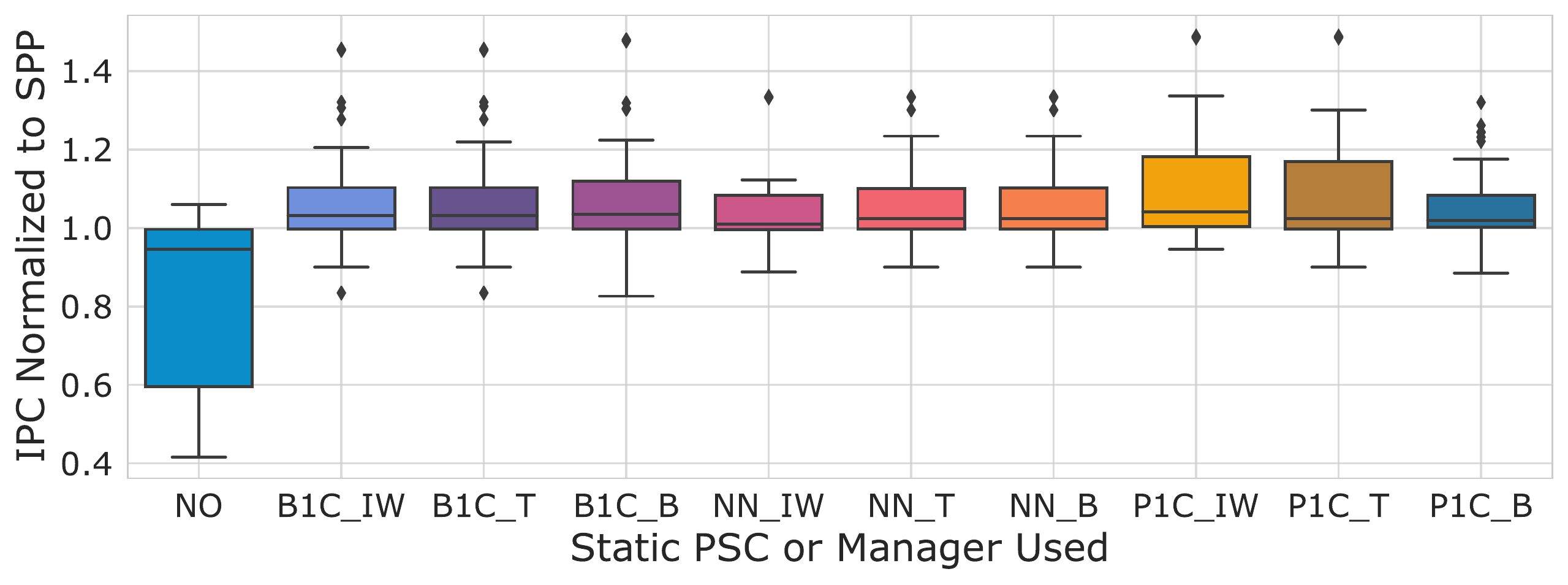}
\caption{\textbf{Training Dataset Study} - Performance distribution of P1C, B1C, and NN normalized to \textit{SPP}, when using three different approaches for generating the training dataset. Here \textit{\_IW} indicates randomly selected instruction windows, \textit{\_T} indicates randomly selected traces, and \textit{\_B} indicates randomly selected benchmarks, for forming the training dataset.}
  \label{fig:1C_unseen}
\vspace{-0.25in}
\end{figure}
}

\section{Evaluation Results}
\label{sec:results}

\subsection{\myalgo~in a 1C Processor}
\label{subsec:1c-analysis}

In this section, we present the processor performance analysis when using \myalgo in 1C, 4C and 8C processors, a sensitivity analysis of how \myalgo's performance varies with cache size, and we explore the trade-off between performance improvement and hardware overhead when using \myalgo. 
We evaluate \myalgo using scope, accuracy, and power metrics. 
For the analysis presented in Sections~\ref{subsec:1c-analysis},~\ref{subsec:4c-analysis},~\ref{subsec:8c-analysis}, and~\ref{subsec:cache-size-sensitivity}, we assume a total hardware overhead budget of 10KB for storing the \myalgo model, and due to its simplicity (as explained in Section~\ref{fig:sysarchitecture}), we assume negligible evaluation overhead for the computing logic.

\begin{figure}[]
  \centering
  \includegraphics[width=0.9\columnwidth]{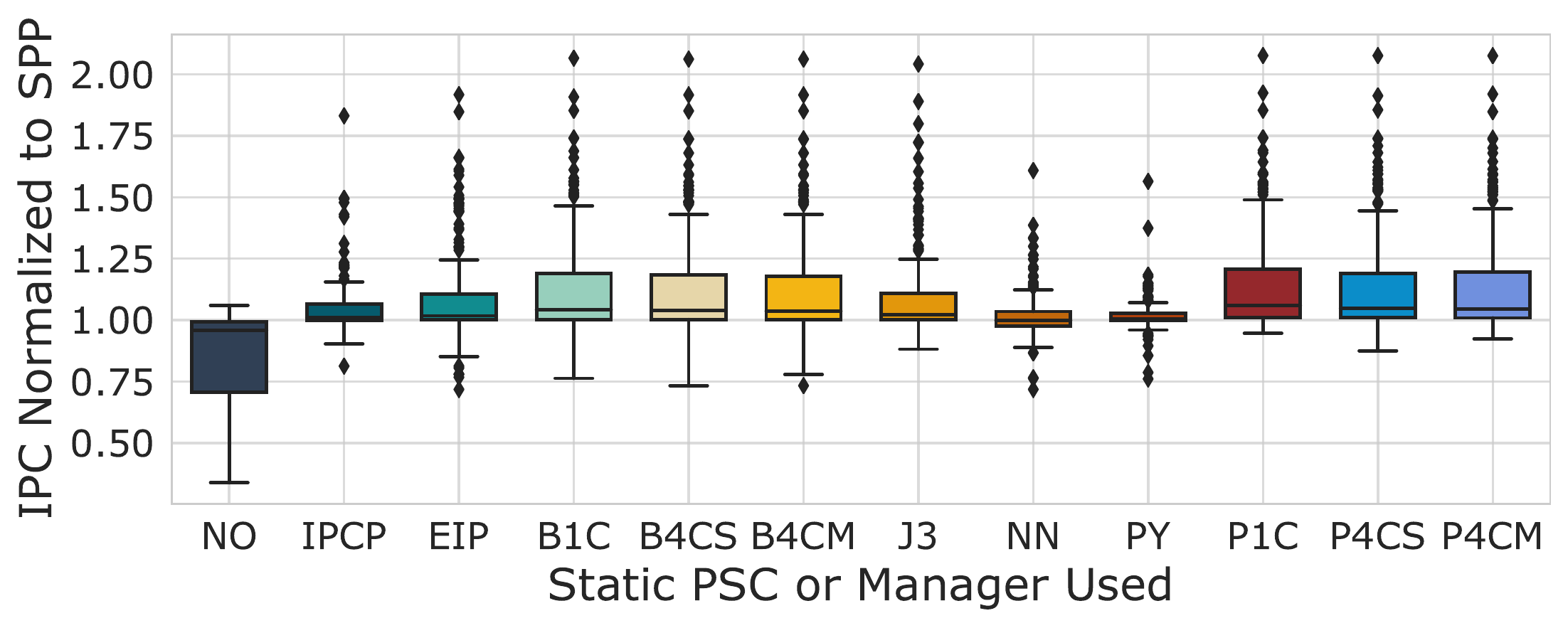}
\caption{\textbf{Normalized Performance of 1C Processor} - Performance distribution of \myalgo~and prior work normalized to \textit{SPP}~\cite{kim2016path}. See Table~\ref{tab:algorithms} for the notations used along the X-axis.}
  \label{fig:1C_box}
\end{figure}

\ignore{
\begin{figure}[]
  \centering
  \includegraphics[width=\columnwidth]{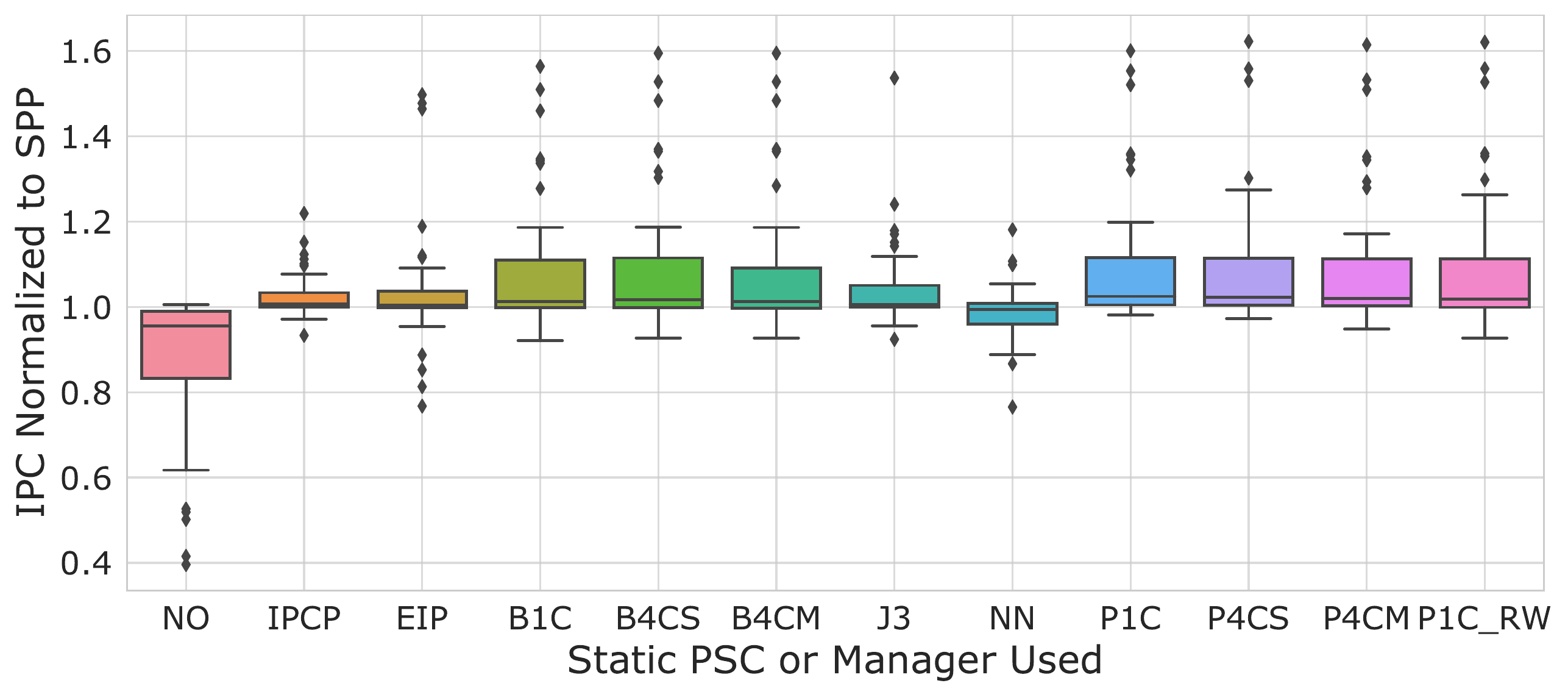}
\caption{\textbf{{\color{blue}Normalized Performance of 1C Processor for Unseen Workloads} - {\color{blue}Performance distribution of \myalgo~and prior work normalized to \textit{SPP}~\cite{kim2016path}. See Table~\ref{tab:algorithms} for the notations used along the X-axis.}}}
  \label{fig:1C_box_RW}
\vspace{-0.1in}
\end{figure}
}

\begin{figure}[]
  \centering
  \includegraphics[width=\columnwidth]{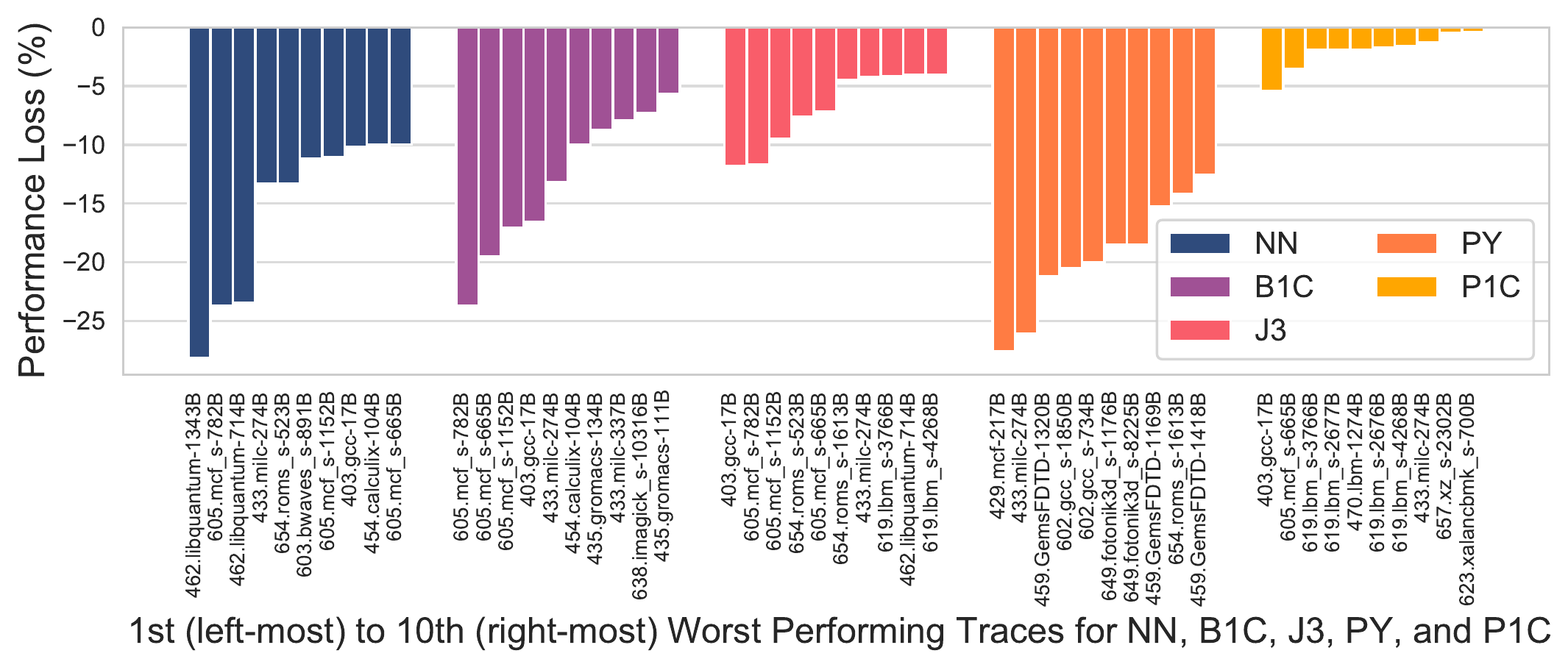}
\caption{\textbf{Bottom Ten Performance Outliers of 1C} - Performance normalized to \textit{SPP}. 
Each group shows the worst performing traces for NN, B1C, J3, and PY, P1C~ordered 10th-worst (right-most) to 1st-worst (left-most).}
  \label{fig:spec_neg}
\vspace{-0.1in}
\end{figure}

{\color{black}
In Figure \ref{fig:1C_box}, we show the IPC distribution for various static PSCs and prefetcher managers that change the PSC at runtime.
We begin our discussion with P1C which is \myalgo~trained using the 1C data suite.
Broadly, compared to a processor with no prefetchers, P1C provides an average performance gain of 46.0\% (peak value of 613\%). 
The true benefit of \myalgo is observed when we look closer into the performance loss in Figure \ref{fig:spec_neg}. 
We observe that when using B1C, 53 traces lose performance and 6 out of those 53 traces lose more than 10\% performance.
This performance loss would make this solution non-acceptable. 
\myalgo~has a worst-case loss of only 5\% and only 8 traces in total have lower performance than SPP. 
This clearly illustrates that \myalgo~provides a win-win situation, whereby we not only see a better average performance gain but also see a reduction in the maximum performance loss and the number of traces that have performance loss.
For the other prior works, we observe that P1C provides 4.65\%, 5.8\%, 12.2\%, 15.3\%, and 5.1\% average IPC gain over IPCP, EIP, NN, PY, and J3, respectively. 

We also trained two more flavors for \myalgo~and BT using 4C STS and 4C MTS datasets. 
P4CS and P4CM achieve 0.5\% lower average performance gain than P1C. 
This is because \myalgo trained using 1C data is better suited for a 1C processor. 
However, the performance improvement when using \myalgo, which is trained with 4C data, is still quite large at 45\% (average of P4CS and P4CM) compared to a system with no prefetching. 
This means that \myalgo~generalizes quite well and has learned the underlying architectural phenomena.
Compared to B1C for B4CS and B4CM, we observe 1\% and 2\% lower average performance, respectively.
Furthermore, the performance loss for the worst case outlier increases to 30\%.
Finally, one thing to note about J3 is that its average IPC gain is 2.1\% lower than B1C, yet the negative outliers are less in both quantity and magnitude. 
This is because J3 makes more conservative changes compared to BT since the algorithm cycles through the PSCs as part of a constantly ongoing trial phase.
This means that, in a production capable system, J3 might have been more viable compared to BT. 
This is because processors have to ensure all applications retain or increase their performance across new processor generations.
}

\subsection{\myalgo~in a 4C Processor}
\label{subsec:4c-analysis}

\begin{figure}[]
  \centering
  \includegraphics[width=0.9\columnwidth]{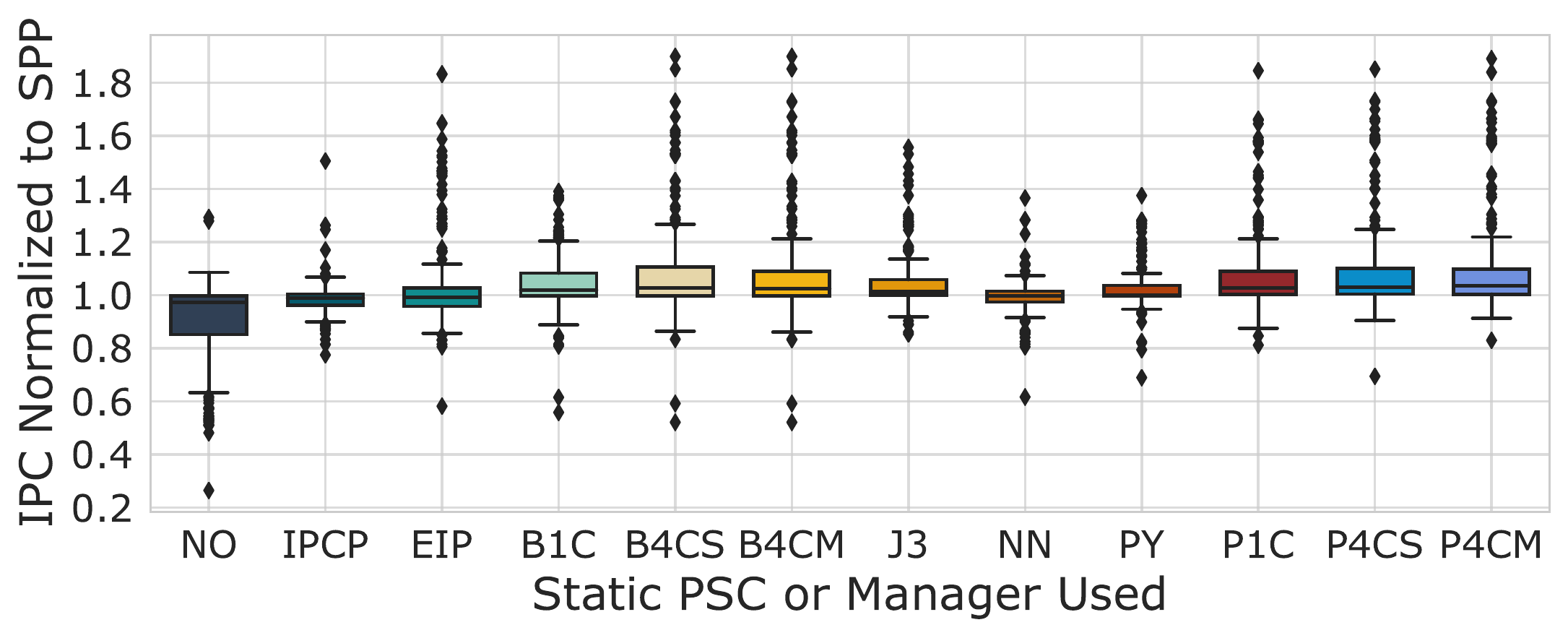}
\vspace{-0.1in}
\caption{\textbf{Normalized Performance of 4C Processor Running STS} - Performance distribution of \myalgo~and prior work normalized to \textit{SPP}. Here the reported performance is average performance across all the cores. See Table~\ref{tab:algorithms} for the notations used along the X-axis.}
  \label{fig:4CVTS_box}
\end{figure}

Here we discuss the use of \myalgo trained using 1C, 4C STS and 4C MTS datasets, in a 4C processor.
These cases are represented by P1C, P4CS and P4CM, respectively.
In Figure~\ref{fig:4CVTS_box}, we show the IPC distribution of \myalgo, and the various prior works in a 4C system while running 4C STS data suite. 
We observe that the average IPC gain of P1C over the no-prefetcher case is 25.8\%.
Compared to B1C, P1C has 4.2\% higher average performance. 
The worst-case performance loss in B1C has gone up to 45\%, while the worst-case loss of P1C is at only 19\%. 
The number of traces that lose performance is 61 for B1C, while P1C has just 24 traces that lose performance.
Compared to IPCP, EIP, NN, PY, and J3, our P1C achieves 10.8\%, 4.8\%, 8.7\%, and 6.8\%, 3.7\% average IPC gain, respectively.
One key observation here is that although P1C has only been trained on 1C data, it is still applicable to the 4C case and provides a clear advantage over prior work.
This is important given that as the number of cores increases, the number of unique combinations of different traces that we will need to run in the multi-core processor increases exponentially (for a 4C processor we need to cover $232^4$ = 2.89 billion trace combinations).
Therefore, generalization using only 1C data is an important aspect to consider when comparing ML-based algorithms.
To further test \myalgo~and BT, we train both algorithms with 4C STS and 4C MTS data suites. 
For \myalgo~we observe that P4CS has 1.2\% and P4CM has 2.2\% better performance compared to P1C (see Figure~\ref{fig:4CVTS_box}). 
While for BT, B4CS has 4.0\% and B4CM has 2.9\% better performance compared to B1C. 
This means that \myalgo~is better at learning the underlying (micro)architectural behavior with a variety of traces running concurrently compared to the same trace running on all the cores.
In contrast BT requires data collected specifically from the same set of experiments to achieve better performance. 

\begin{figure}[]
  \centering
  \includegraphics[width=0.9\columnwidth]{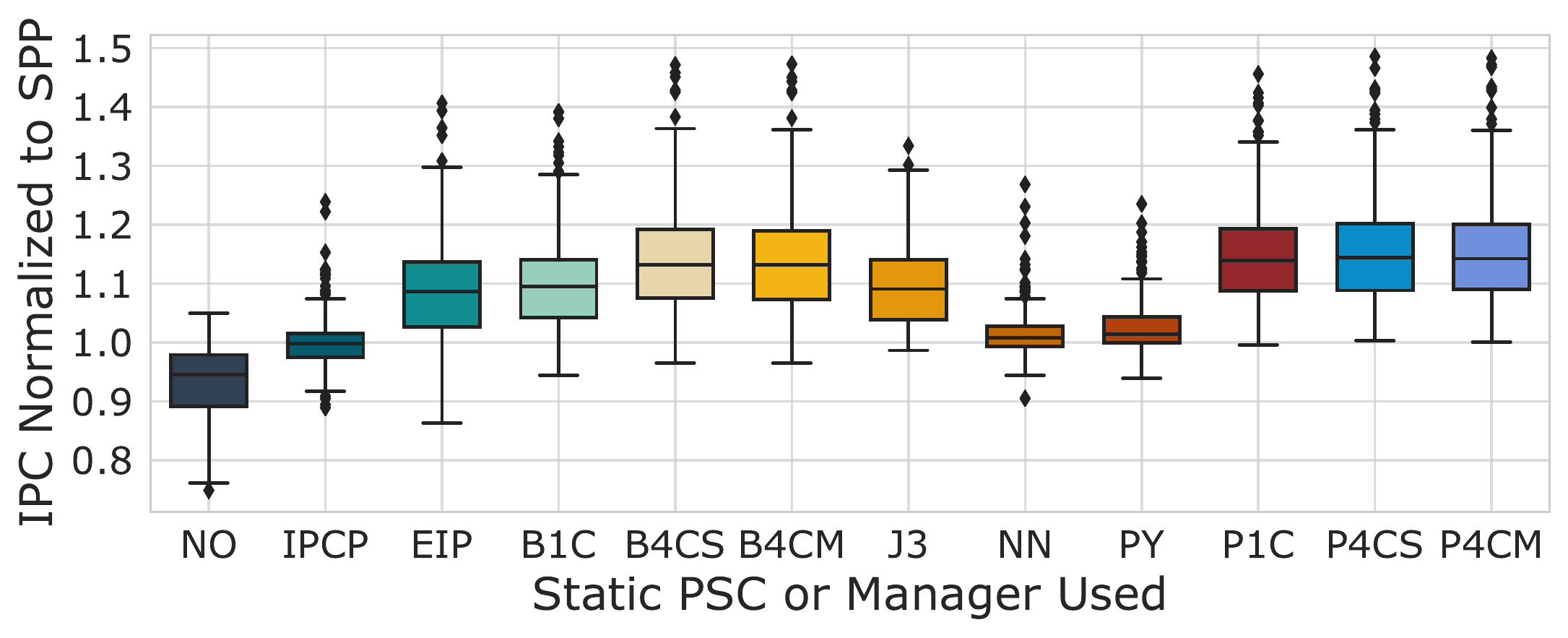}
  \vspace{-0.1in}
\caption{\textbf{Normalized Performance of 4C Processor Running MTS} - Performance distribution of \myalgo~and prior work normalized to \textit{SPP}. Here the reported performance is average performance across all the cores. See Table~\ref{tab:algorithms} for the notations used along the X-axis.}
  \label{fig:4CMTS_box}
  \vspace{-0.1in}
\end{figure}

In Figure \ref{fig:4CMTS_box} we show the IPC distribution of \myalgo~and the various prior work when running 4C MTS. 
P1C achieves 23\% average performance gain over no prefetching. 
P1C's average performance gain is also very close ($<$0.6\% difference) to the average performance gain of P4CS and P4CM. 
Once again, we observe that \myalgo~is superior to BT in this regard with B1C achieving 5\% lower performance gain than P1C. 
B1C's performance gain is also 4\% lower than B4CS and 4.5\% lower than B4CM. 
When we observe the outliers, B1C has 23 outliers while P1C has only 3. 
While the worst case outlier in B1C has 7\% performance loss compared to SPP, P1C has only 0.4\% loss.
Compared to IPCP, EIP, NN, and PY, J3, our P1C (and also P4CS and P4CM) achieves 14.5\%, 5\%, 12.8\%, and 11.9\%, 4.4\% average IPC gain, respectively. 

\subsection{\myalgo~in an 8C Processor}
\label{subsec:8c-analysis}

\begin{figure}[t]
  \centering
  \includegraphics[width=0.9\columnwidth]{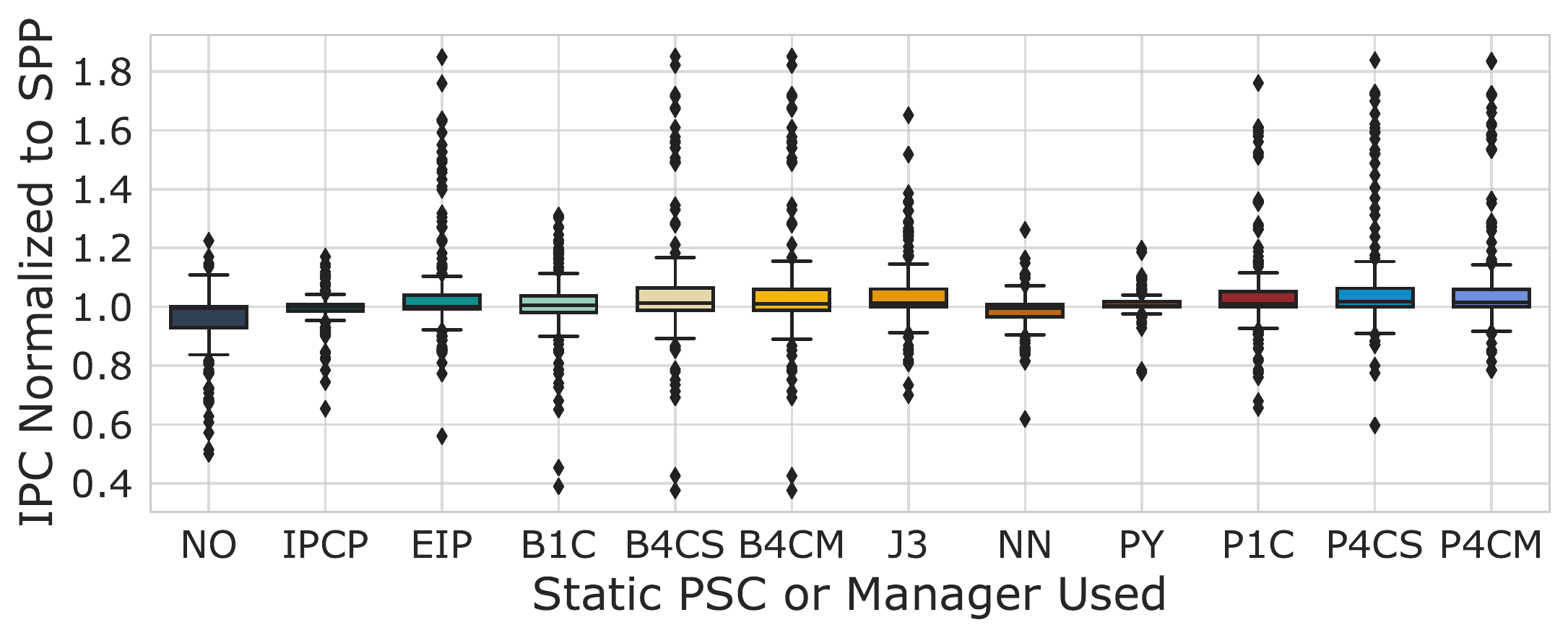}
\vspace{-0.1in}
\caption{\textbf{Normalized Performance of 8C Processor Running STS} - Performance distribution of \myalgo~and prior work normalized to \textit{SPP}. Here the reported performance is average performance across all the cores. See Table~\ref{tab:algorithms} for the notations used along the X-axis.}
  \label{fig:8CVTS_box}
\end{figure}

\begin{figure}[]
  \centering
  \includegraphics[width=0.9\columnwidth]{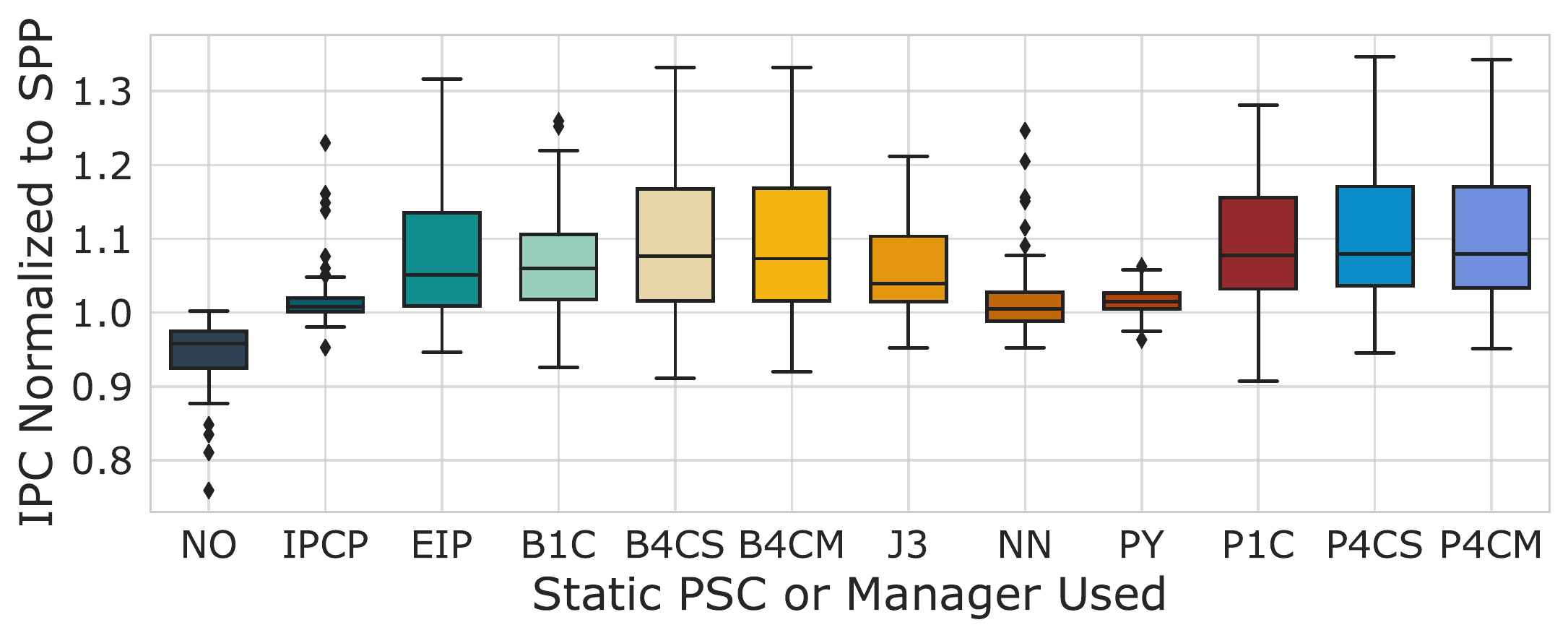}
\vspace{-0.1in}
\caption{\textbf{Normalized Performance of 8C Processor Running MTS} - Performance distribution of \myalgo~and prior work normalized to \textit{SPP}. Here the reported performance is average performance across all the cores. See Table~\ref{tab:algorithms} for the notations used along the X-axis.}
  \label{fig:8CMTS_box}
\vspace{-0.1in}
\end{figure}

Here we discuss the use of \myalgo trained using 1C, 4C STS and 4C MTS datasets, in an 8C processor.
These cases are represented by P1C, P4CS and P4CM, respectively.
We conduct this 8C processor analysis to test if \myalgo~scales to a larger number of cores. For an 8C processor running STS, we observe several interesting trends (see Figure~\ref{fig:8CVTS_box}). 
P1C has 11.9\% average IPC gain over no prefetching, which is 4.8\% higher than B1C.
With P4CS and P4CM our IPC gain is 12.7\% and 12.9\%, respectively, over the no prefetching case. 
This means that to train \myalgo~for a multicore processor, we should have at least some data corresponding to a multicore processor in our dataset.
It should be noted that, compared to the no prefetching case, P1C, P4CS and P4CM have lower IPC gain in the 8C processor than the 4C processor, which in turn has lower performance gain than the 1C processor.
This is because prefetching is much harder to do in multi-core processors. 
The overall benefit of prefetching goes down in multi-core processors and \myalgo~has lower possible peak performance.
Comparatively, B1C has almost 0\% performance gain over SPP.
Even a static PSC, namely EIP, has 5.5\% higher average IPC compared to B1C.
P1C also gives comparable performance to B4CS and B4CM even though P1C was not trained using any data from a multicore processor.
When we observe the outliers, B1C has 83 outliers with worst case performance loss of 62\%; while, P1C has 47 outliers with worst case performance loss of 35\%. 
Compared to IPCP, EIP, NN, PY, and J3, our P1C achieves 5.9\%, 0.2\%, 6.5\%, and 4.4\%, 1.2\% average IPC gain, respectively. 

When using the 8C MTS (see Figure~\ref{fig:8CMTS_box}), we observe similar trends to when using 8C STS. 
P1C has 2\% better average IPC than B1C and comparable performance to B4CS and B4CM.
P4CS and P4CM achieve 4\% and 4.4\% better average IPC compared to B1C.
Compared to IPCP, EIP, NN, and PY, J3, our P1C achieves 6.9\%, 1.6\%, 6.6\%, 7.6\%, and 3.1\% average IPC gain, respectively.

\subsection{Puppeteer Performance vs Different Cache Sizes}
\label{subsec:cache-size-sensitivity}

\begin{figure}[]
  \centering
  \includegraphics[width=0.8\columnwidth]{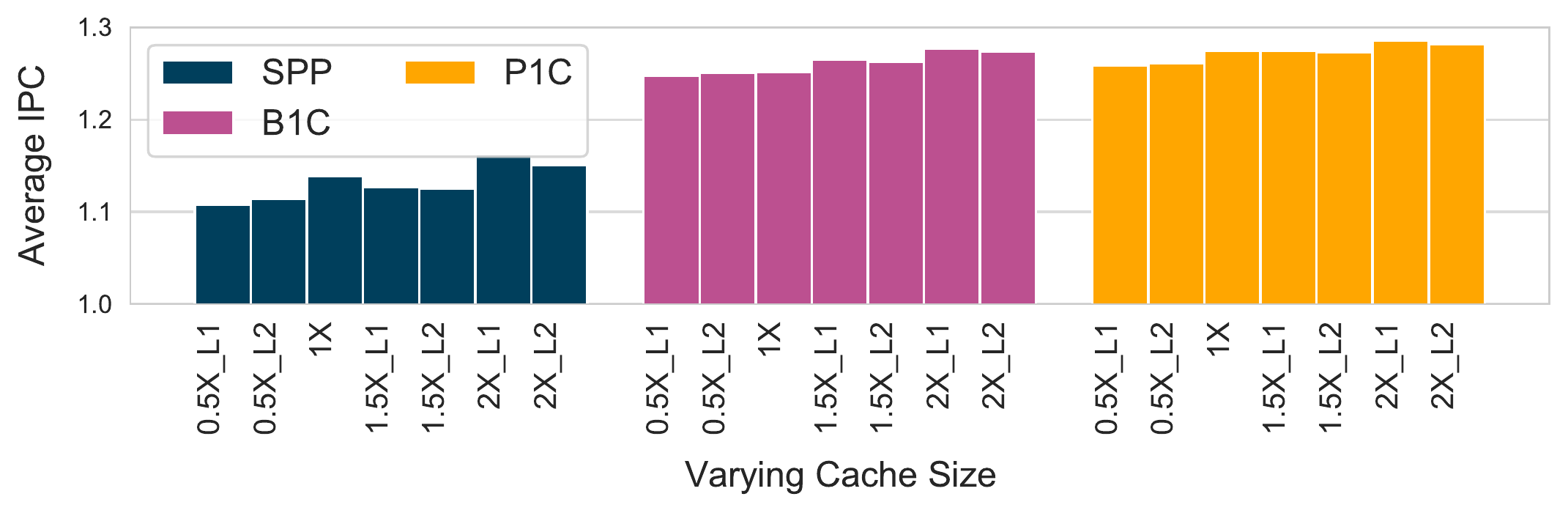}
\vspace{-0.1in}
\caption{\textbf{Cache Size Sensitivity Study} - Average performance of B1C, SPP, and P1C for $0.5\times$, $1\times$, $1.5\times$, and $2\times$ the nominal L1\$ and L2\$ sizes. For each bar we change only the L1\$ size or the L2\$ size.}
  \label{fig:cache_study}
\vspace{-0.15in}
\end{figure}

In this subsection we discuss how the performance of BT, SPP and \myalgo varies with cache size.
We train both models only using data collected on a 1C processor with 32KB L1I\$, 48KB L1D\$, 512KB L2\$, and 2MB LLC.
In Figure~\ref{fig:cache_study} we show the average IPC values of BT (i.e. B1C), SPP, and \myalgo (i.e. P1C) for $0.5\times$, $1\times$, $1.5\times$, and $2\times$ the nominal L1\$ and L2\$ sizes.
P1C is affected by L1\$ size slightly more than L2\$ size but the difference is small (at most 0.4\% more performance at $2\times$ L1\$ compared to $2\times$ L2\$). 
At $0.5\times$ L1\$ size P1C has 1.5\% lower performance compared to the performance of P1C at nominal cache size. 
At $2\times$ L1\$ size P1C has 1.1\% better performance compared to the performance of P1C at nominal cache size. At all cache sizes, P1C performs at least 1\% better than B1C with the largest performance difference of 2.3\% at nominal cache size.
SPP performance swings by 5.5\% between the $2\times$ L1\$ and the $0.5\times$ L1\$, and by 3.7\% between the $2\times$ L2\$ and the $0.5\times$ L1\$. 
In contrast, P1C has only a 2.9\% swing. 
This means that with P1C we are already operating close to the possible peak performance and so increasing the cache size does not have a significant effect on the performance.

\subsection{Puppeteer Performance vs Different Hardware Overheads}
\label{subsec:overhead-sensitivity}

\begin{figure}[b]
  \centering
\vspace{-0.3in}
  \includegraphics[width=\columnwidth]{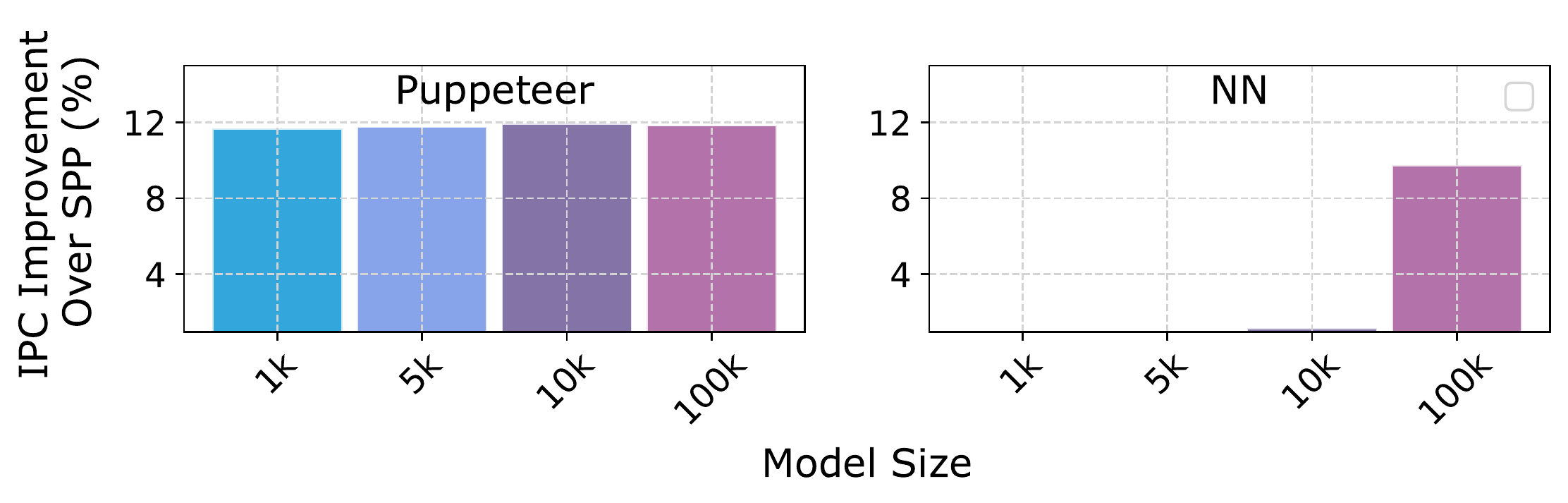}
\vspace{-0.1in}
\caption{\textbf{Model Size Scaling} - Average IPC Improvement on 1C Data Suite of \myalgo~and NN for Different Model Sizes.}
  \label{fig:model_size}
\end{figure}

\begin{figure*}[]
  \centering
  \includegraphics[width=0.9\textwidth]{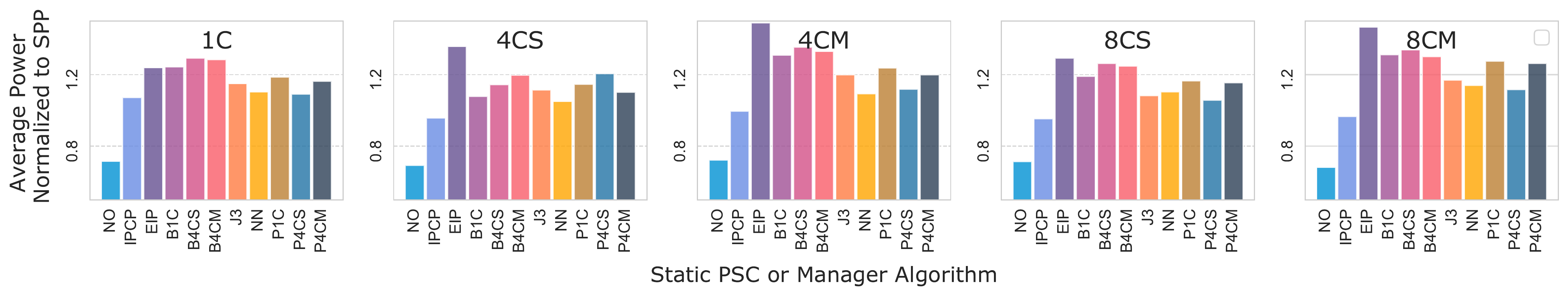}
  \vspace{-0.1in}
\caption{\textbf{Average power of \myalgo~and prior work normalized to \textit{SPP}:} Here 1C = 1C data suite, 4CS = 4C single-type trace set data suite, 4CM = 4C mixed-type trace set data suite, 8CS = 8C single-type trace set data suite, and 8CM = 8C mixed-type trace set data suite.}
  \label{fig:power}
  \vspace{-0.3in}
\end{figure*}

In this subsection we study the performance of \myalgo and NN when we have four different hardware overhead budgets -- 1KB, 5KB, 10KB, and 100KB. 
Here we train a different \myalgo model and a NN model for each hardware overhead budget.
In Table \ref{tab:algo_power_area} we show the total area and power required for the NN and \myalgo.
All designs in Table \ref{tab:algo_power_area} are such that they need less than 1\% of the instruction widow time to choose a PSC.
We calculate the power, area, and delay values using Cadence Genus and SRAM array compiler for 22 nm GLOBALFOUNDRIES\textsuperscript{\textregistered} (GF22FDX) \cite{carter201622nm}.
\ignore{In Table \ref{tab:components} we show the various logic and memory components that are required for \myalgo~and NN for the different models.
We calculate the power, area, and delay values of the components using Cadence Genus and SRAM array compiler for 22 nm GLOBALFOUNDRIES (GF22FDX)\textsuperscript{\textregistered} \cite{carter201622nm}.}

To select a PSC, \myalgo~just traverses through the random forest for each PSC.
{\color{black}If we evaluate all 5 forests in series, where we will require a maximum 250 comparison operations (5 forests $\times$ 5 trees per forest $\times$ (10 comparisons for 1KB, 5KB, and 10KB; 20 comparisons for 100KB) = 250 or 500 comparisons),} it will take less than 0.5\% of the total time required to execute the 100K instructions in the instruction window (assuming each instruction takes on average a clock cycle).
Thus, we end up using the chosen PSC for 99.5\% of the instruction window for \myalgo.
In contrast, the NN will require a multiply and addition operation for each weight, as well as a division and ReLu function at the end of each layer to determine the PSC.
Due to the heavy computation required, as the size of the NN increases, to choose a PSC in less than 1\% of the instruction window time, NN will need to parallelize the operations, which will require additional logic and memory components.

\begin{table}[t]
  \centering
  \vspace{0.1in}
  \caption{\textbf{Power and Area for various sizes of \myalgo~and NN:} Here we use SRAM compiler for designing Node MEM, and design the compute logic using RTL and then synthesize it using Cadence Genus for GF22FDX\textsuperscript{\textregistered} \cite{carter201622nm}. Power is given over one instruction window. \textit{Norm.} values are w.r.t. to \myalgo~1KB values.}
  \vspace{-0.1in}
  \label{tab:algo_power_area}
\scriptsize
\renewcommand{\arraystretch}{1.0}
\centering
\center{
\begin{tabular}{c|c|c|c|c|c}
\hline
\hline
\multicolumn{2}{c|}{\textbf{Algorithm}} & \textbf{Area (\SI{}{\micro\metre ^2})} & \textbf{Area Norm.} & \textbf{Power ((\SI{}{\micro W}))} & \textbf{Power Norm.} \\
\hline
\multicolumn{2}{c|}{\myalgo 1KB} &  5,000 &  $1\times$ & 4.7 & $1\times$\\
\hline
\multicolumn{2}{c|}{\myalgo 5KB} &  10,000 &  $2\times$ & 8 & $1.7\times$\\
\hline
\multicolumn{2}{c|}{\myalgo 10KB} &  18,000 &  $3.6\times$ & 9 & $1.9\times$\\
\hline
\multicolumn{2}{c|}{\myalgo 100KB} &  140,000 &  $28\times$ & 39 & $8.3\times$\\
\hline
\multicolumn{2}{c|}{NN 1KB} &  17,000 &  $3.4\times$ & 14 & $3\times$\\
\hline
\multicolumn{2}{c|}{NN 5KB} &  40,000 &  $8\times$ & 65 & $13.8\times$\\
\hline
\multicolumn{2}{c|}{NN 10KB} &  55,000 &  $11\times$ & 100 & $21\times$\\
\hline
\multicolumn{2}{c|}{NN 100KB} &  380,000 &  $76\times$ & 870 & $185\times$\\
\hline
\hline
\end{tabular}
\vspace{-0.1in}
}
\end{table}

In Figure \ref{fig:model_size} we show the average IPC for 1C for the differently sized models.
As it is immediately apparent, \myalgo provides good performance improvement even when using a small model that fits within 1KB, while NN does not provide any performance improvement until we use a model that requires 10KB. 
At 10KB model size, a 1C processor with a NN manager has 11.8\% lower average IPC as compared to a 1C-processor with \myalgo.
The NN also has $\sim 3\times$ larger area and $11\times$ larger power as compared to \myalgo.
If we compare 100KB NN to 1KB \myalgo, then using \myalgo provides 11.6\% performance improvement, while NN provides 9.7\% performance improvement while the area and power of 100KB NN is $76\times$ and $185\times$, respectively, that of \myalgo. 
Therefore, NN is not suitable for hardware prefetcher adaptation. 

\subsection{Puppeteer Power Analysis}

In Figure \ref{fig:power} we show the average power consumed in the caches for various static PSCs and the prior managers compared to \myalgo.
Compared to the no prefetching case, overall \myalgo has 65\% higher power consumption.
Compared to BT~(average of B1C, B4CS, and B4CM), \myalgo~(average of P1C, P4CS, and P4CM)~has 9\% lower average power. 
While NN and J3 have 6.3\% and 2.2\% lower average power than \myalgo, as we have discussed they also have significantly lower average IPC. 
For the static PSC, EIP is extremely power hungry, with the highest average power that is 20.4\% more than \myalgo. 
IPCP and NO have lower power consumption than \myalgo~but they also have the worst performance among all options we have considered.

\subsection{Puppeteer's impact on Accuracy and Scope}

\begin{figure}[t]
  \centering
  \includegraphics[width=0.9\columnwidth]{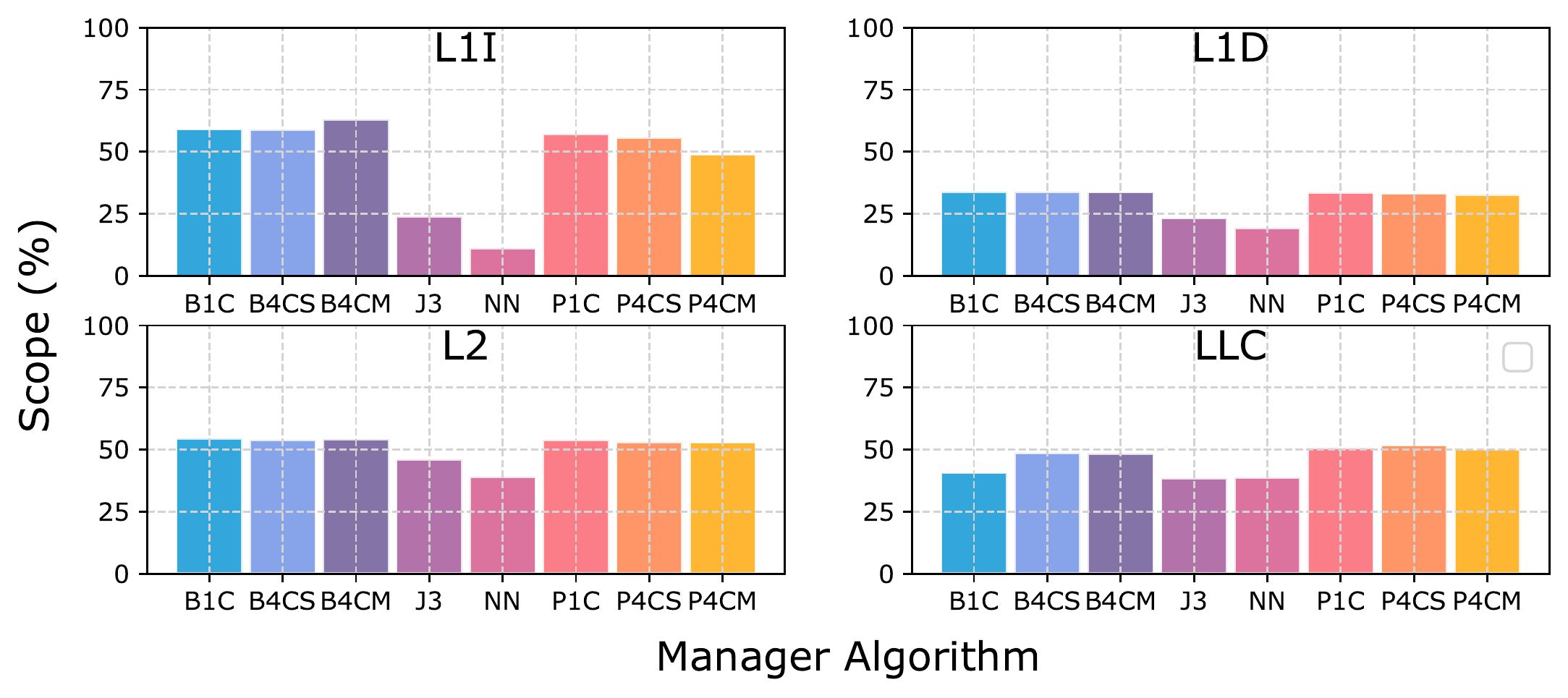}
\vspace{-0.1in}
\caption{\textbf{Average scope of \myalgo~and the prior works in a 1C processor:} See Table~\ref{tab:algorithms} for the notations used along the X-axis.}
  \label{fig:cov}
  \vspace{-0.2in}
\end{figure}

In Figure \ref{fig:cov} we compare the prefetching scope of the prior works and \myalgo. 
Here we determine scope as the number of total misses reduced by the prefetcher or prefetching system, divided by the total number of misses with prefetching disabled.
We observe very limited scope across the 4 levels of the memory hierarchy for NN and J3.
This is probably one of the reasons these options have lower performance than BT and \myalgo.
Comparing BT~(average of B1C, B4CS, and B4CM), \myalgo~(average of P1C, P4CS, and P4CM), in L1D\$ and L2\$ BT achieves a marginal difference with 0.5\% broader scope than \myalgo.
In L1I\$ cache BT actually has 3.3\% broader scope compared to \myalgo.
However, \myalgo~has 4.8\% broader scope compared to BT in LLC. 
Intuitively, since LLC penalties are more important than L1I\$, L1D\$, and L2\$ penalties, this is a better outcome.
Experimentally, our results also support this outcome since \myalgo~has better performance than BT.

\begin{figure}[t]
  \centering
  \includegraphics[width=0.9\columnwidth]{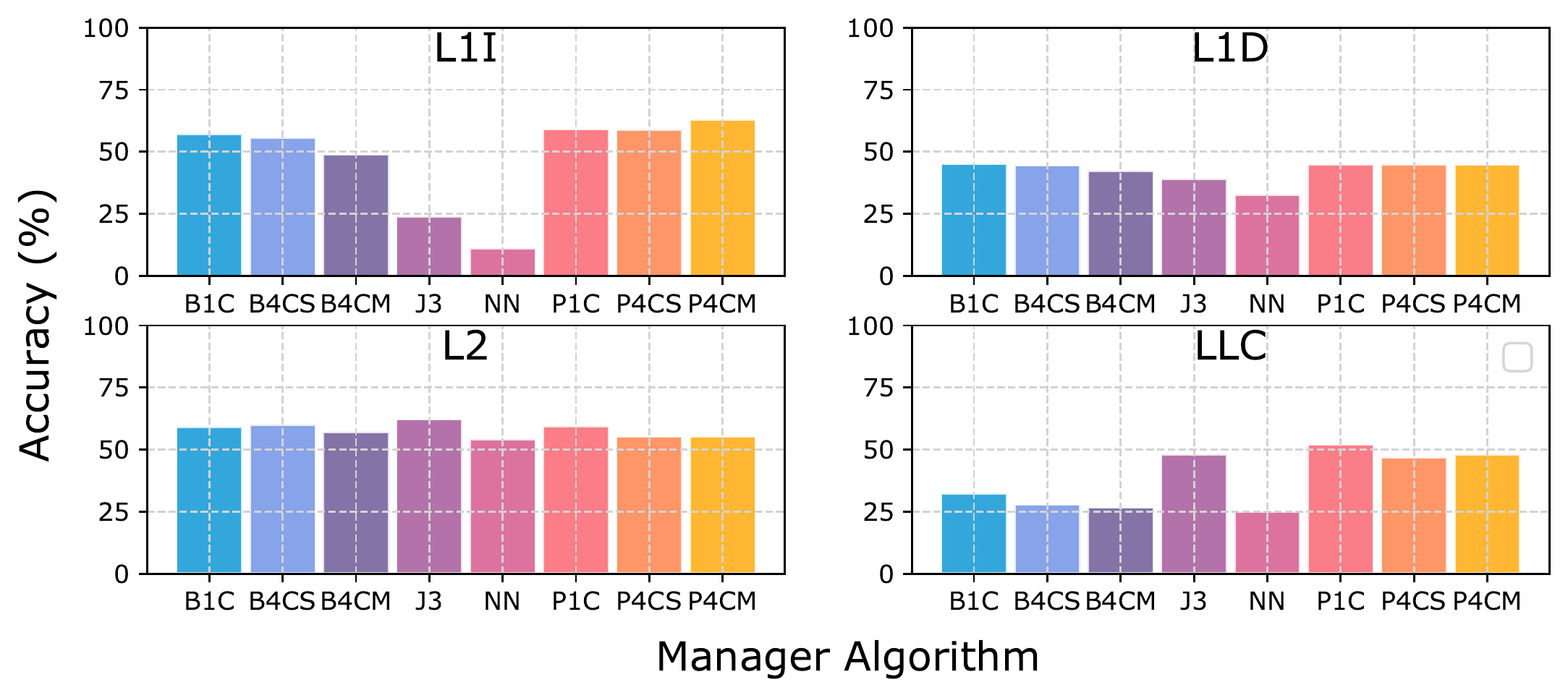}
\vspace{-0.1in}
\caption{\textbf{Average prefetching accuracy of \myalgo~and the prior works in a 1C processor:} See Table~\ref{tab:algorithms} for the notations used along the X-axis.}
  \label{fig:acc}
\vspace{-0.2in}
\end{figure}

In Figure \ref{fig:acc}, we compare the prefetching accuracy of \myalgo~and the prior works. 
Here accuracy of a prefetcher is measured as the number of misses that a prefetcher or prefetcher system has reduced compared to the case when prefetching is disabled, divided by the number of misses caused by the prefetcher. 
\myalgo~is better than the other options in L1I\$ and L1D\$, but comparable in L2\$. 
In LLC, J3 fairs better than the other options with only 1.5\% lower accuracy than \myalgo.
In case of BT and NN, \myalgo achieves 21\% better accuracy compared to BT and 28\% better accuracy compared to NN.
It is also of note that since the accuracy of \myalgo~in L1I\$ is 6.7\% better compared to BT, this better accuracy plays a role in compensating for the broader L1I\$ scope of BT compared to \myalgo. 

\subsection{\myalgo-based Temporal variations in PSC}

\begin{figure}[t]
  \centering
    \subfloat[PSC Percentage]{
    \includegraphics[clip,width=0.27\columnwidth]{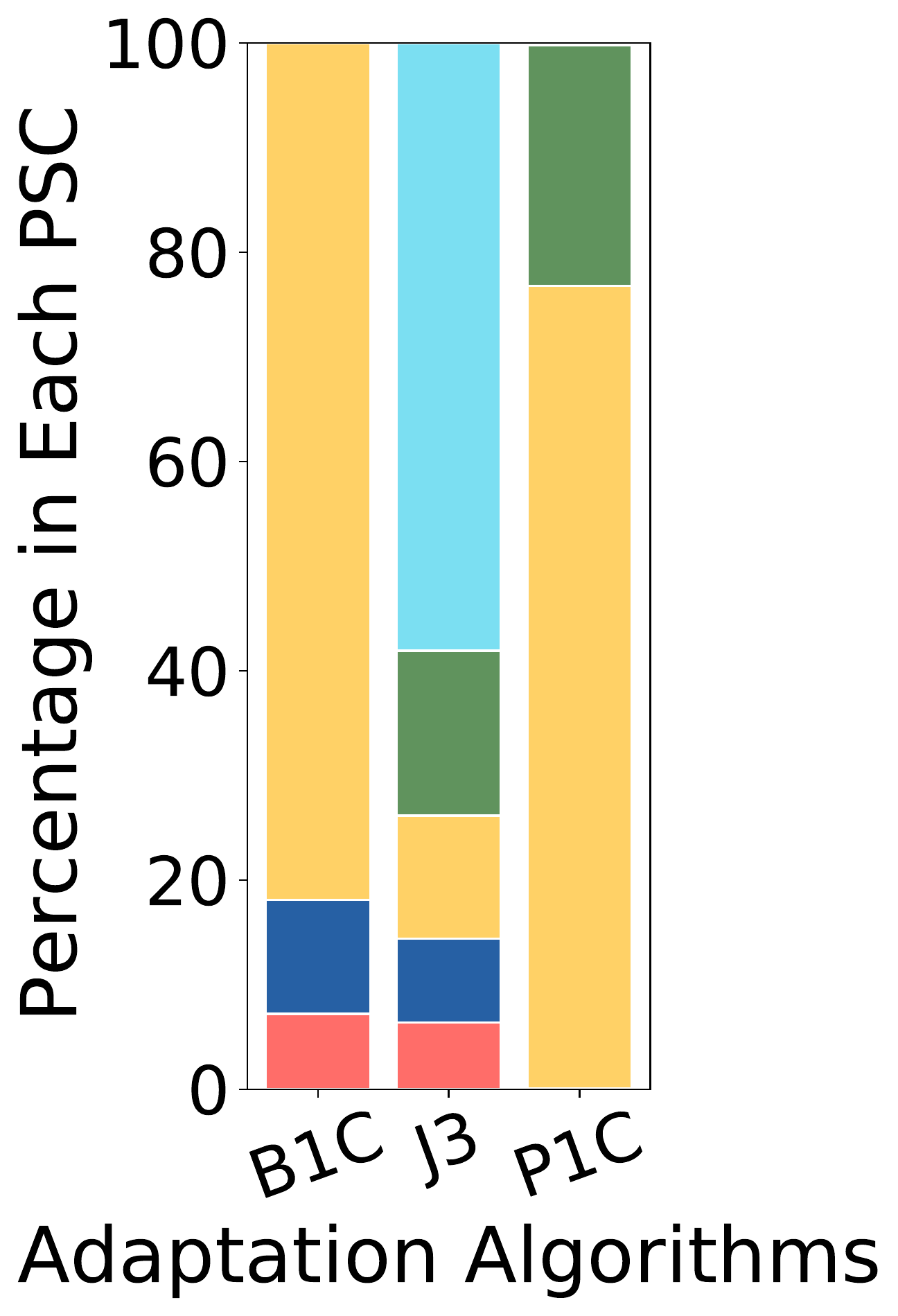}
  } 
  \subfloat[Runtime Behavior]{
    \includegraphics[clip,width=0.67\columnwidth]{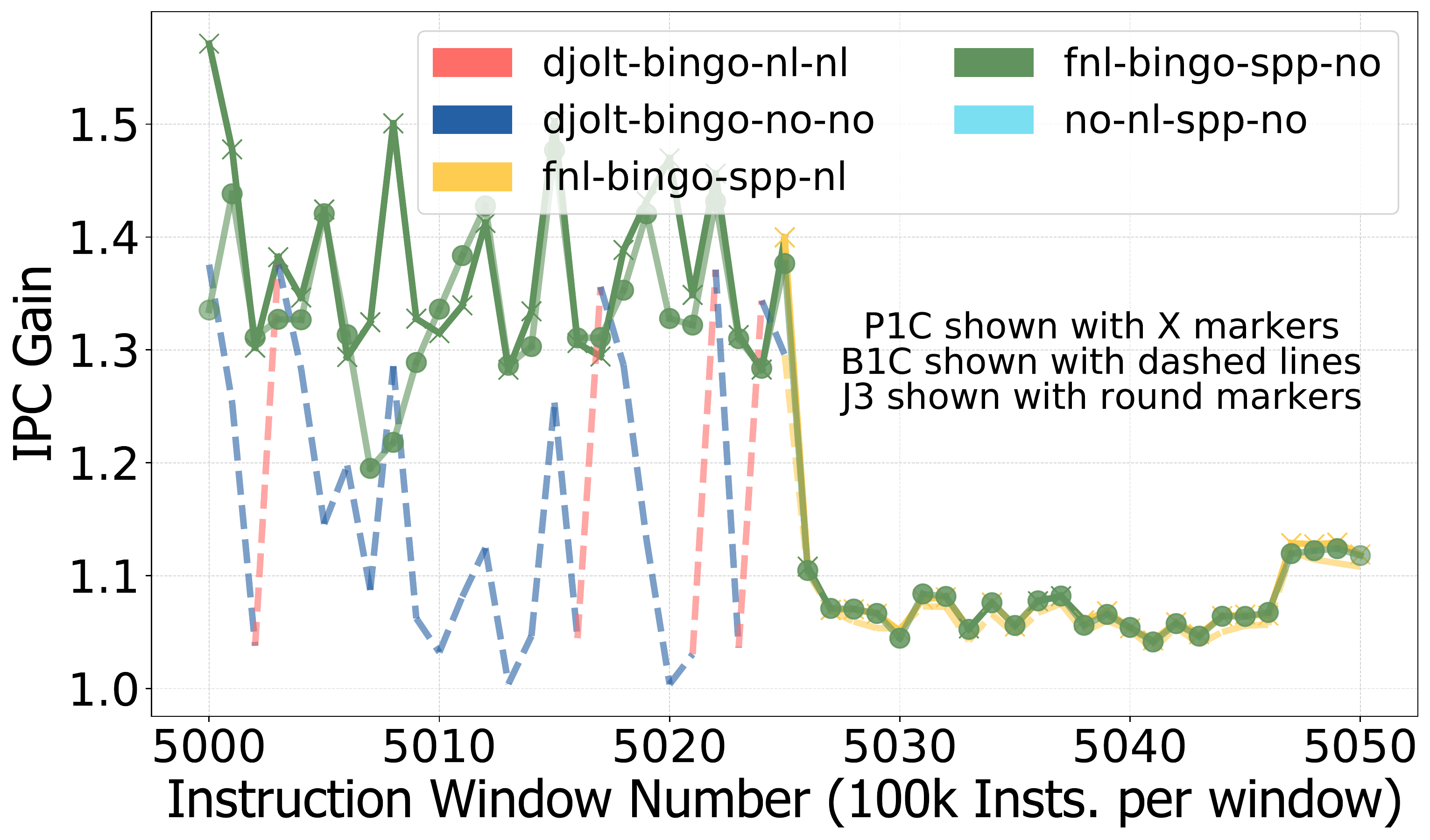}
  }
  \caption{\textbf{Temporal Behavior - } (a)  Percentage usage of each PSC for the P1C, B1C, and J3; and (b) IPC gain when using P1C, B1C and J3 across 50 instruction windows and running 429.mcf-217B. For each plot line, for an instruction window we use color coding to indicate the PSC choice. The PSC descriptions are provided in Table \ref{tab:final_psc}.}
  \label{fig:temporal}
\vspace{-0.2in}
\end{figure}

In Figure \ref{fig:temporal} we show the temporal behavior of P1C, B1C, and J3 while running 429.mcf-217B as an example. 
Figure \ref{fig:temporal}a shows the percentage of time for which each PSC was used by each manager algorithm when executing 1.2B instructions of 429.mcf-217B.
In Figure \ref{fig:temporal}b we show the IPC values and the PSC used in each instruction window for a small slice of the same trace. 
We would like to note three interesting observations:
(i) From Figure~\ref{fig:temporal}a, both B1C and P1C use \textit{fnl-bingo-spp-no} for the 80\% of the instruction windows, yet the performance difference between the two is around 20\% over the whole trace.
This means the PSC chosen in the remaining 20\% of the instruction windows have a larger influence on the overall performance.
(ii) In the first 25 instruction windows after the 5000th instruction window, there is a large variation in IPC gain when using B1C as compared to P1C, while all three algorithms converge to the same performance and same PSC during the last 25 instruction windows.
This shows that P1C does a better job at predicting the PSC in different regions of an application.
(iii) J3 uses the same PSC as P1C yet has lower performance in the first 25 instruction windows. 
This illustrates that changing the PSC has a cumulative effect on IPC. 
The choice of PSC made by P1C in prior instruction windows allowed P1C to gain more performance in the given instruction windows compared to J3. 

\ignore{{\color{red}TODO: Some quantification of the percentage/amount of time \myalgo~is better than other algorithms.}}
\section{Conclusion and Future Work} 
In this work, we introduce \myalgo, a novel ML-based prefetcher manager designed using custom tailored random forests.
We train a dedicated random forest for each PSC, which allows the random forest to retain more information in a smaller amount of hardware. 
For the 232 traces that we evaluated, \myalgo~achieves an average performance gain of 46.0\% in 1C, 25.8\% in 4C, and 11.9\% in 8C compared to a system with no prefetching. 
{\color{black}\myalgo~also reduces the number of negative outliers by 89\%.}
As future work, we will explore a unified design of a ML-based manager that selects from an array of ML-based prefetchers. 
In addition, we will also consider online training to further improve the performance of Puppeteer.

\newpage
\bibliographystyle{IEEEtranS}
\bibliography{puppeteer-isca-2022.bib}

\end{document}